\documentclass[a4paper,12pt]{article}
\usepackage[utf8]{inputenc}
\usepackage{amsmath}  
\usepackage{amssymb}       
\usepackage{amsfonts}
\usepackage{bbm}
\usepackage{times}
\usepackage{amsthm} 
\usepackage{epsfig}
\usepackage{paralist}         
\usepackage{color}
\newcommand{\GREEN}[1]{\textcolor{green}{#1}} 
\newcommand{\RED}[1]{#1} 
\newcommand{\BLUE}[1]{#1} 
\newcommand{\sout}[1]{} 
\newcommand{\RRRED}[1]{#1} 
\newcommand{\BBBLUE}[1]{#1} 
\newcommand{\sssout}[1]{} 
\newcommand{\DATUM}{19.05.2022}       
\newcommand{\ol}{\overline}   
  
\newcommand{\eps}{{\varepsilon}}  
\newcommand{\vphi}{{\varphi}}           
\newcommand{\Om}{\Omega}                
\newcommand{\om}{\omega}
\newcommand{\la}{\langle}
\newcommand{\ra}{\rangle}
\newcommand{\udarrow}{\uparrow, \downarrow}
\newcommand{\bfone}{\mathbf{1}}
\newcommand{\cA}{\mathcal{A}}
\newcommand{\cB}{\mathcal{B}}
\newcommand{\cD}{\mathcal{D}}
\newcommand{\cE}{\mathcal{E}}

\newcommand{\cL}{\mathcal{L}}         
         
\newcommand{\cN}{\mathcal{N}}         
\newcommand{\cO}{\mathcal{O}}         
\newcommand{\cP}{\mathcal{P}}

\newcommand{\cS}{\mathcal{S}}
\newcommand{\cT}{\mathcal{T}}
\newcommand{\cU}{\mathcal{U}}

\newcommand{\CC}{\mathbbm{C}}     
\newcommand{\EE}{\mathbbm{E}} 
\newcommand{\HH}{\mathbbm{H}} 
\newcommand{\tHH}{{\widetilde{\mathbbm{H}}}} 
\newcommand{\hh}{\mathbbm{h}}     
\newcommand{\NN}{\mathbbm{N}}     
\newcommand{\PP}{\mathbbm{P}}     
\newcommand{\bbp}{\mathbbm{p}}     
\newcommand{\vpp}{{\vec{\mathbbm{p}}}}     
\newcommand{\QQ}{\mathbbm{Q}}     
\newcommand{\RR}{\mathbbm{R}}     
\newcommand{\UU}{\mathbbm{U}} 
\newcommand{\tUU}{{\widetilde{\mathbbm{U}}}} 
     
\newcommand{\VV}{\mathbbm{V}}     
     
\newcommand{\ZZ}{\mathbbm{Z}}     
\newcommand{\fF}{\mathfrak{F}}
\newcommand{\fG}{\mathfrak{G}}  
\newcommand{\fH}{\mathfrak{H}}

\newcommand{\fh}{\mathfrak{h}}  

\newcommand{\hfh}{{\hat{\mathfrak{h}}}}
 
\newcommand{\fg}{\mathfrak{g}} 
\newcommand{\fs}{\mathfrak{s}}  
\newcommand{\sfj}{\mathsf{j}}
\newcommand{\hsfj}{\hat{\mathsf{j}}}
\newcommand{\sfJ}{\mathsf{J}}

\newcommand{\uR}{{\underline R}} 

\newcommand{\uZ}{{\underline Z}}

\newcommand{\uf}{{\underline f}}
\newcommand{\ug}{{\underline g}}

\newcommand{\ux}{{\underline x}}

\newcommand{\uz}{{\underline z}}

\newcommand{\unu}{{\underline \nu}}

\newcommand{\hE}{\widehat{E}}
\newcommand{\hcE}{\widehat{\mathcal{E}}}

\newcommand{\hT}{\widehat{T}}    
\newcommand{\hV}{\widehat{V}}

\newcommand{\halpha}{\hat{\alpha}}
\newcommand{\hGamma}{\widehat{\Gamma}}
\newcommand{\hgamma}{{\hat{\gamma}}}

\newcommand{\tE}{\widetilde{E}}

\newcommand{\tF}{\widetilde{F}}
\newcommand{\tH}{\widetilde{H}}

\newcommand{\tGamma}{\widetilde{\Gamma}}

\newcommand{\tlambda}{\tilde{\lambda}}

\newcommand{\trho}{\widetilde{\rho}}

\newcommand{\vR}{{\vec{R}}}

\newcommand{\vsigma}{{\vec{\sigma}}}
\newcommand{\vnabla}{{\vec{\nabla}}}

\newcommand{\va}{{\vec{a}}}

\newcommand{\vv}{{\vec{v}}}
\newcommand{\vx}{{\vec{x}}}
\newcommand{\vy}{{\vec{y}}}

\newcommand{\Bog}{{\mathrm{Bog}}}

\newcommand{\DM}{\mathfrak{DM}}

\newcommand{\QDM}{\mathfrak{QDM}}

\newcommand{\fin}{\mathrm{fin}}

\newcommand{\cirS}{\mathop{\bigcirc\kern -.73em {\scriptstyle{\rm S}}}}

\newcommand{\huelle}{{\rm span}}

\newcommand{\Tr}{{\rm Tr}}

\newcommand{\gs}{{\rm gs}}
\newcommand{\HF}{{\rm HF}}
\newcommand{\TF}{{\rm TF}}
\newcommand{\GPQ}{{\rm GPQ}}
\newcommand{\GPQT}{{\rm GPQ:T}}
\newcommand{\BHF}{{\BLUE{\rm BHF}}}
\newcommand{\BCS}{{\rm BCS}}

\newcommand{\trial}{{\mathrm{trial}}}
\newcommand{\Ex}{{\mathrm{Ex}}}
\newcommand{\aux}{{\mathrm{aux}}}
\newcommand{\av}{{\mathrm{av}}}

\newcommand{\para}{{\mathrm{para}}}
\newcommand{\nonpos}{{\BLUE{\mathrm{neg}}}}
\newcommand{\genonepdm}{\fG^{(1)}}
\newcommand{\onepdm}{\fg^{(1)}}
\newtheorem{theorem}{Theorem}
\newtheorem{lemma}[theorem]{Lemma}            
\newtheorem{corollary}[theorem]{Corollary}

\theoremstyle{plain}
\begin{document}
\bibliographystyle{plain}
\title{Hartree--Fock Theory, \\ Lieb's Variational Principle, 
\\ and their Generalizations}

\author{Volker~Bach $<$v.bach@tu-bs.de$>$, \\[1ex]
Institut für Analysis und Algebra \\
TU Braunschweig \\ Universitätsplatz~2 \\
38106 Braunschweig \\ Germany}

\date{\DATUM}

\maketitle

\begin{abstract}
\noindent \BLUE{Abstract: Hartree--Fock theory in quantum mechanics is
reviewed, from the proposal of the Hartree--Fock approximation right
after quantum mechanics was formulated to its applications in modern
physics. This includes the description of traditional Hartree--Fock
theory in quantum chemistry, its generalizations of various kinds, and
its importance for predicting the presence of symmetry breaking, or
the absence thereof.}
\end{abstract}
%


\centerline{\textbf{Dedicated to Elliott H.\ Lieb}}

\vspace*{3ex}



\thispagestyle{empty}

\newpage
\setcounter{page}{1}

\newpage
\section{Hartree-Fock Theory of Coulomb Systems} \label{sec-I}
%
One of the biggest triumphes of twentieth century science has been the
discovery of quantum mechanics (and quantum field theory) almost one
hundred years ago by Heisenberg \cite{Heisenberg1925}, Born,
Heisenberg, and Jordan \cite{BornHeisenbergJordan1926}, Schr\"odinger
\cite{Schroedinger1926}, Dirac \cite{Dirac1927, Dirac1928a,
  Dirac1928b}, and Pauli \cite{Pauli1928}. It is a remarkable fact
that, while quantum mechanics is of key importance for all
technologies discovered in the past century and the complexity of
theoretic descriptions of quantum systems has increased by several
orders of magnitude, the basic conceptual framework of a complex
Hilbert space $\fH$ of wave functions $\psi(t) \in \fH$ which
represent physical states at time $t \in \RR$ and evolve according to
the (time-dependent) Schr\"odinger equation $i \dot{\psi}(t) = H
\psi(t)$, \BLUE{or observables $A(t)$ which evolve according to the
  Heisenberg equation of motion $\dot{A}(t) = i[H, A(t)]$,} with 
$H = H^*$ being the self-adjoint Hamiltonian operator,
\BLUE{is \hspace*{-5mm} =} \RED{remains} unchanged until today. 
In \RED{the} absence of external fields, the Hamiltonian $H$ is
independent of time $t$. Then the solution of the time-dependent
Schr\"odinger equation can be traced back to determining the spectral
resolution, in particular, all eigenvalues $E \in \RR$ and all
corresponding eigenvectors $\psi_E \in \fH$ of $H$.  Eigenvalues and
corresponding eigenvectors do not cover all possible cases, and the
general task \BLUE{of} \RED{determining} the spectral resolution of
$H$ has lead to the mathematical theory of spectral analysis of
self-adjoint operators, see~\cite{ReedSimonI-IV1980}.

For a Coulomb system, i.e., a nonrelativistic atom ($K=1$) or molecule
($K \geq 2$), consisting of $N \in\ZZ^+ := \{1, 2, 3, \ldots\}$
dynamical, mutually repelling electrons revolving about $K \in \ZZ^+$
attractive static nuclei of charges 
$\uZ := (Z_1, Z_2, \ldots, Z_K) \in [\RR_0^+]^K$ at pairwise distinct
positions $\uR := (\vR_1, \vR_2, \ldots, \vR_K) \in [\RR^3]^K$, the
Hilbert space of a single electron is 
$\fh := L^2( \RR^3 \times \{\uparrow, \downarrow\})$, and the Hilbert
space
\begin{align} \label{eq-I.01}
\fH^{(N)}
\ := \ 
\Big\{ \Psi \in 
\fh^{\otimes N} \; \Big|
\ \forall \, \pi \in S_N: 
\Psi(\ux_\pi) \: = \: (-1)^\pi \, \Psi(\ux) \Big\} 
\end{align}
of the wave function of the system of $N$ dynamical electrons is 
the space of square-integrable functions of $N$ coordinates
$x_n = (\vx_n, \tau_n) \in \RR^3 \times \{\uparrow, \downarrow\}$
which are antisymmetric under
permutations $\ux_{\, \pi} = (x_{\pi(1)}, \ldots, x_{\pi(N)})$ of these
coordinates $(x_1, \ldots, x_N)$. (Here and henceforth we follow the
convention from physics and assume for Hilbert \RED{spaces} that
$(\vphi, \psi) \mapsto \la \vphi | \psi \ra$ is \RED{alway} 
antilinear in $\vphi$ and linear in $\psi$.) The Hamiltonian generating the
dynamics of these $N$ electrons is 
\begin{align} \label{eq-I.02}
H_N(\uZ,\uR) 
\ := \  
\sum_{n=1}^N 
\bigg\{ -\Delta_n - \sum_{k=1}^K \frac{Z_k}{|\vx_n-\vR_k|} \bigg\}
\; + \; \sum_{1 \leq m < n \leq N} \frac{1}{|\vx_m-\vx_n|} \, .
\end{align}
Note that the charges $\uZ$ and the positions $\uR$ of the nuclei
enter $H_N$ as fixed parameters, and $H_N(\uZ,\uR)$ may be considered the
Born--Oppenheimer approximation \cite{BornOppenheimer1927} to lowest
order. \BLUE{Frequently we do not} display the dependence of the Hamiltonian
on $\uZ$ and $\uR$ and simply write $H_N \equiv H_N(\uZ,\uR)$. The
Hamiltonian $H_N$ is essentially self-adjoint and semibounded on the
space $\cS^{\wedge N} := \cS^{\otimes N} \cap \fH^{(N)}$ of
antisymmetric Schwartz test functions of $N$ variables which is a
dense subspace of $\fH^{(N)}$. (We henceforth largely ignore domain
questions, \RED{do not display} $\cS^{\wedge N}$, and implicitly assume
sufficient regularity of the wave functions under considertation.)

The semiboundedness of $H_N$ ensures the finiteness of the
\RED{\textit{ground state energy}}, i.e., the infimum $\RED{E_\gs(N)}
\equiv \RED{E_\gs(N,\uZ,\uR)} := \inf\sigma( H_N )$ of the
spectrum of $H_N$. The ground state energy $E_\gs(N)$ and, if
$E_\gs(N)$ happens to be an eigenvalue, the corresponding
\RED{\textit{ground state}} (eigenvector) $\RED{\Psi_\gs} \in
\fH^{(N)}$ are basic quantities for the physical description of the
Coulomb system. The actual solution of the corresponding eigenvalue
equation $H_N \Psi_\gs = E_\gs(N) \Psi_\gs$, however, is inaccessible
to explicit solution or even numerical computation for large molecules
due to the large number of variables involved.

At this point the Rayleigh--Ritz principle becomes of key importance
because it yields a variational characterization of
\begin{align} \label{eq-I.03}
E_\gs(N) 
\ = \  
\inf\Big\{ \la \Psi \, | \; H_N \Psi \ra \ \Big| \
\Psi \in \fH^{(N)} \, , \ \|\Psi\| = 1 \Big\} 
\end{align}
as the lowest energy expectation value the Hamiltonian $H_N$ admits.
Instead of solving the Schr\"odinger equation -which is virtually
impossible- one computes the energy expectation value of any
normalized trial state 
$\Psi_\trial \in \fH^{(N)}$. This yields an upper
bound $\la \Psi_\trial | H_N \Psi_\trial \ra \geq E_\gs(N)$ on the
ground state energy. If 
$\la \Psi_\trial | H_N \Psi_\trial \ra - E_\gs(N)$ is small, the trial
state $\Psi_\trial$ is assumed to be a good approximation to (one of)
the actual ground state(s) $\Psi_\gs$. The mathematical justification
for this replacement, e.g., in terms of quantitative error bounds, is
a difficult and largely open mathematical problem.

The earliest and, perhaps, most natural choice of trial states for
Coulomb systems made is known as the \RED{\textit{Hartree--Fock
  approximation}}\BBBLUE{, which had been originally proposed by Hartree
  \cite{Hartree1928} but without an antisymmetry contraint on the wave
  function. This was followed by improvements} \RED{by} Fock \cite{Fock1930}
and Slater \cite{Slater1930, Slater1951}\BBBLUE{, who} took the
antisymmetry of the trial state correctly into account. It is a
variational principle in which the variation in \eqref{eq-I.03} is
restricted to \RED{\textit{Slater determinants}}, i.e., to wave functions of
the form $\RED{\Phi(\uf)} := f_1 \wedge f_2 \wedge \cdots
\wedge f_N$.  These are antisymmetrized tensor products
\begin{align} \label{eq-I.04}
f_1 \wedge f_2 \wedge \cdots \wedge f_N
\ := \  
\frac{1}{\sqrt{N!}} \, \sum_{\pi \in \cS_N} (-1)^\pi \:
f_{\pi(1)} \otimes f_{\pi(2)} \otimes \cdots \otimes f_{\pi(N)} 
\end{align}
of $N$-tuples $\uf = ( f_1, \ldots , f_N ) \in \fh^N$ of mutually
orthonormal \RED{\textit{orbitals}}, i.e., vectors $f_i \in \fh$ in the
\RED{\textit{one-particle Hilbert space}}
\begin{align} \label{eq-I.05}
\RED{\fh} 
\ = \ 
L^2\big( \RR^3 \times \{\uparrow, \downarrow\} \big) \, ,
\end{align}
obeying $\la f_i | f_j \ra_\fh = \delta_{i,j}$ The
corresponding infimum
\begin{align} \label{eq-I.06}
\RED{E_\HF(N)}
\ := \  
\inf\Big\{ \big\la \Phi(\uf) \, \big| \; H_N \, \Phi(\uf) \big\ra 
\ \Big| \ f_1, \ldots , f_N \in \: \fh
\, , \ \la f_i | f_j \ra_\fh = \delta_{i,j} \Big\} 
\end{align}
is called the \RED{\textit{Hartree--Fock \BBBLUE{ground state} energy}}. A
straightforward computation gives
\begin{align} \label{eq-I.07}
\cE_\HF(\uf) \ := \ &
\big\la \Phi(\uf) \, \big| \, H_N \Phi(\uf) \big\ra 
\\[1ex] \nonumber 
\ = \ &
\sum_{i=1}^N \la f_i \, | \, h f_i \ra_\fh 
\; + \; \frac{1}{2} \sum_{i,j=1}^N 
\la f_i \wedge f_j \, | \, V (f_i \wedge f_j) \ra_{\fh \otimes \fh} \, ,
\end{align}
where the \RED{\textit{one-particle operator}} 
$\RED{h} := -\Delta - \sum_{k=1}^K Z_k \, |\vx-\vR_k|^{-1}$ is
a second-order differential operator acting on (a suitable dense
domain in) $\fh$, and the \RED{\textit{pair interaction potential}}
$\RED{V} := |\vx-\vy|^{-1}$ is a multiplication operator on (a
dense domain in) $\fh \otimes \fh$. The energy functional
$\cE_\HF(\uf)$ can be written as a sum 
\begin{align} \label{eq-I.07,01}
\cE_\HF(\uf)
\ = \ &
T(\uf) \: - \: U(\rho_{\uf}) 
\: + \: \frac{1}{2} D(\rho_{\uf}) 
\: - \: \frac{1}{2} X(\gamma_{\uf}) 
\end{align}
of the \RED{\textit{kinetic energy $\RED{T(\uf)}$}}
minus the \RED{\textit{nuclear attraction $\RED{U(\rho_{\uf})}$}},
\begin{align} \label{eq-I.08}
\BBBLUE{T}(\uf)
\ := \ 
\sum_{\tau = \udarrow} \GREEN{\sum_{i=1}^N} \int |\vnabla f_i( \vx, \tau)|^2 \: d^3x 
\, , \quad
\BBBLUE{U}(\rho)
\ := \ 
\sum_{k=1}^K \int \frac{Z_k \, \rho(\vx)}{|\vx-\vR_k|} \: d^3x \, , 
\end{align}
plus the \RED{\textit{direct term 
\BBBLUE{$\RED{\frac{1}{2} D(\rho_{\uf})}$}}},
representing the classical electrostatic energy, minus the
\BBBLUE{\RED{\textit{exchange term 
$\RED{\frac{1}{2} X(\gamma_{\uf})}$}}},
\begin{align} \label{eq-I.09}
\BBBLUE{D(\rho)}
\, :=  
\iint \frac{\rho(\vx) \, \rho(\vy) \; d^3x \, d^3y}{|\vx-\vy|} 
\, , \quad 
\BBBLUE{X(\gamma)}
\, := 
\iint \frac{|\gamma(\vx, \vy)|^2  \; d^3x \, d^3y}{|\vx-\vy|} \, ,
\end{align}
where $\gamma_{\uf}(\vx, \vy) := \sum_{\tau = \udarrow} \sum_{i=1}^N
f_i(\vx,\tau) \, \ol{f_i(\vy,\tau)}$ and 
$\rho_{\uf}(\vx) := \gamma_{\uf}(\vx, \vx)$ are the one-particle
density matrix and the one-particle density corresponding to
$\Phi(\uf)$, respectively. The explicit and relatively simple \RED{forms}
of these terms are one main reason for the success of the
Hartree--Fock approximation. 

For large neutral Coulomb systems, i.e., for $Z = N \gg 1$,
$\uZ = Z \uz$, with $\uz = (z_1, \ldots, z_k)$ for fixed $z_k >0$ 
\RED{summing to $1$}, and nuclear positions 
$\uR(Z) = \big( \vR_1(Z), \ldots, \vR_k(Z) \big)$ not too
close to each other, 
$\inf_{Z >0} \min_{k< \ell} 
\{ Z^{1/3} |\BLUE{\vR_k(Z)} - \BLUE{\vR_\ell(Z)}| \} > 0$,
the Hartree--Fock energy is seen \cite{Bach1992} to obey
\begin{align} \label{eq-I.09,01}
E_\HF(Z) \ = \ 
E_\TF(Z, Z \uz, \uR(Z)) \: + \: \frac{Z^2}{4} \sum_{k=1}^K z_k^2 
\: + \: \cO\big( Z^{5/3} \big) \, ,
\end{align}
where the main contribution to leading order in $Z$ is the
\RED{\textit{Thomas--Fermi energy $\RED{E_\TF\big(Z, Z \uz,
    \uR(Z)\big)}$}} established by Lieb and Simon in
\cite{LiebSimon1977a}, which is bounded above and below by universal
multiples of $Z^{7/3}$, \RED{and} followed by the \RED{\textit{Scott correction
  $\RED{\frac{Z^2}{4} \sum_{k=1}^K z_k^2}$}} of order $Z^2$
derived by Hughes \cite{Hughes1986} and by Siedentop and Weikard
\cite{SiedentopWeikard1987a, SiedentopWeikard1989a} for atoms ($K=1$),
by Ivrii and Sigal \cite{IvriiSigal1993} for molecules 
($K \geq 1$) and by \RED{Solovej, S{\o}rensen, and Spitzer} in the relativistic
and nonrelativistic case for both atoms and molecules 
\cite{SolovejSpitzer2003, SolovejSoerensenSpitzer2010}. \BBBLUE{(See
also \RED{Siedentop's} contribution to this volume.)}

For our discussion we observe that if $\uf^{(\HF)}$ is a
minimizer of $\cE_\HF$ (or an \textit{approximate minimizer},
i.e., $\cE_\HF(\uf^{(\HF)}) \leq E_\HF(N) + \eps$, for $\eps >0$ sufficiently
small) under the
orthonormality constraint $\la f_i | f_j \ra_\fh = \delta_{i,j}$, then
there exist universal constants $0 < c < C < \infty$ such that, for
any choice of $\uz = (z_1, \ldots, z_k)$ and 
$\uR(Z) = \big( \vR_1(Z), \ldots, \vR_k(Z) \big)$,
\begin{align} \label{eq-I.09,02}
c \, Z^{7/3} 
\ \leq \ 
T(\uf^{(\HF)}) \, , \ U(\rho_\HF) & \, , \ D(\rho_\HF)
\ \leq \  
C \, Z^{7/3} \, , 
\\[1ex] \label{eq-I.09,03}
c \, Z^{5/3} 
\ \leq \ 
X(\gamma_\HF) & 
\ \leq \  
C \, Z^{5/3} \, ,
\end{align}
where $\rho_\HF := \rho_{\uf^{(\HF)}}$ and 
$\gamma_\HF := \gamma_{\uf^{(\HF)}}$. 
That is, the kinetic energy, the nuclear attraction, and the classical
electrostatic energy are all or the order $Z^{7/3}$, while the
exchange energy is of order $Z^{5/3}$ and hence much smaller in magnitude.

The dominance of the three contributions $T(\uf^{(\HF)})$,
$U(\rho_\HF)$, and $D(\rho_\HF)$ to the
energy compared to the contribution of the exchange term
$X(\gamma_\HF)$ can be anticipated from the
Cauchy-Schwarz inequality which implies that 
\BLUE{$X(\gamma_\uf) \leq D(\rho_\uf)$},
for any model with repulsive pair interaction 
$V(x-y) \geq 0$.  Note, however, that this takes only \BLUE{total}
ground state energies of the entire system into account; if we compare
energy \BLUE{differences}, then the exchange contribution may become
the decisive quantity that determines whether a system
binds or not.

Furthermore, if 
$\uf^{(\HF)}$ is a minimizer of $\cE_\HF$ under the orthonormality
constraint $\la f_i | f_j \ra_\fh = \delta_{i,j}$ then $\cE_\HF$ is
stationary at $\uf^{(\HF)} = (f_1^{(\HF)}, \ldots, f_N^{(\HF)})$ and
the Euler--Lagrange equations -known in this context as
\RED{\textit{Hartree--Fock equations}}- become
\begin{align} \label{eq-I.10}
h_\HF[\uf^{(\HF)}] \, f_i^{(\HF)} \ = \ e_i \, f_i^{(\HF)} \, ,
\end{align}
for all $i \in \{1, \ldots, N\}$, where the eigenvalues $e_i$
are Lagrange multipliers imposed to fulfill the orthonormality constraint
and $\RED{h_\HF[\uf^{(\HF)}]}$ is the 
\RED{\textit{Hartree--Fock effective Hamiltonian}}
acting on orbitals $g \in \fh$ as
\begin{align} \label{eq-I.11}
\big( h_\HF[\uf] & \, g \big)[\vx,\tau] \ := \ 
\\[1ex] \nonumber 
& \big( h \, g \big)[\vx,\tau] \: + \:
\bigg( \int \frac{\rho_{\uf}(\vy) \, d^3y}{|\vx-\vy|} \bigg) g(\vx,\tau)
\: - \:
\int \frac{\gamma_{\uf}(\vx, \vy) \, g(\vy,\tau) \: d^3y}{|\vx-\vy|} \, .
\end{align}
Even though the Hartree--Fock equations form a system
of nonlinear partial integro-differential equations in $\uf$, the
reduction of $N$ dynamical variables to \RED{$2$} makes it accessible to
numerical solution. Concrete numerical algorithms to solve the
Hartree--Fock equations have been analyzed mathematically by Cancès
and Le~Bris in \cite{CancesLeBris2000}. More recently, the numerical
solution of the corresponding self-consistent equation of generalized
Hartree--Fock theory described in Section~\ref{sec-VI}, the
\GREEN{Bogoliubov}--Hartree--Fock equations, has been studied by Lewin and Paul
in \cite{LewinPaul2014}.

\BBBLUE{Hartree--Fock theory is closely related to density functional
theory and \textit{Kohn--Sham (KS) theory}. These latter two are based
on the Hohenberg--Kohn theorem \cite{HohenbergKohn1964} which
asserts that the ground state energy of any Coulomb system can be
expressed as the infimum of a universal (but unknown) functional 
of the electron density only. A mathematically precise formulation
of the Hohenberg--Kohn theorem was given by Levy \cite{Levy1979} and 
Lieb \cite{Lieb1983}. We describe the Kohn--Sham theory from the 
viewpoint of Hartree--Fock theory, although this oversimplifies
their physical arguments somewhat. Namely, Kohn and Sham proposed 
to approximate the exchange term \BLUE{$X(\gamma_\uf)$} by a 
functional \BLUE{$\int G[\rho_\uf(\vx)] \, d^3x$} 
of the one-particle density \BLUE{$\rho_\uf$} only, where $G$ is yet to be
determined. A natural candidate for $G$ is $G[\rho] = 
C_{\text{Dirac}} \, \rho^{4/3}$, which \RED{was} proposed by Dirac
in \cite{Dirac1930} and whose quality as an approximation to
the exchange term \RED{was} analyzed in \cite{Bach1993}. 
This approximation is known as the \textit{local density approximation
  (LDA)} and the \textit{KS-LDA} theory is widely and successfully
used in numerical studies in material science. Its mathematical 
foundation including a proof of existence of minimizers
of the Kohn--Sham energy functional and, hence, of solutions
of the corresponding stationarity condition known as \RED{the} 
Kohn--Sham equations
was given by Anantharaman and Cancès in \cite{AnantharamanCances2009}.
An important improvement to the local density approximation defined by a 
function $G[\rho]$ is the \textit{generalized gradient approximation (GGA)}. 
It accommodates an additional dependence of the exchange term function 
\BLUE{$G[\rho, \nabla \sqrt{\rho}]$} on the gradient of the (square root
of the) density, leading to KS-GGA theory. A very successful proposal
for the form of \BLUE{$G[\rho, \nabla \sqrt{\rho}]$} was made by
Perdew, Burke, and Ernzerhof in \cite{PerdewBurkeErnzerhof1996}
and is known as \textit{PBE}. }

The first mathematically rigorous treatment of the Hartree--Fock
approximation and the corresponding Hartree--Fock equations was given
by Lieb and Simon in \cite{LiebSimon1977b}. By applying the
so-called direct methods of the calculus of variations, they prove the
existence of a minimizer $\uf^{(\HF)}$ of $\cE_\HF$, which then
necessarly fulfills the Hartree--Fock equations, under the condition
that the number $N-1$ of electrons minus one is strictly less than the
total nuclear charge $Z := \sum_{k=1}^K Z_k$. This is a natural
HVZ-\BLUE{(Hunziker-van Winter-Zhislin-)}type condition reflecting the fact that, if one electron is
spatially separated far away from the nuclei, it is still attracted by
a Coulomb force induced by a net charge $Z-N+1 >0$. This force binds
this outer electron to the molecule and prevents its escape to
infinity.

An important technical point in \cite{LiebSimon1977b} is the
conversion of the original orthonormality condition
$\la f_i | f_j \ra_\fh = \delta_{i,j}$ into the equivalent statement
$G(\uf , \uf) = \bfone$ on $\CC^N$, where 
$G(\uf , \ug)_{i,j} := \la f_i | g_j \ra_\fh$ denotes the Gram matrix 
of $\uf = (f_1, \ldots, f_N), \ug = (g_1, \ldots, g_N) \in \fh^N$.
Lieb and Simon then observe that the minimization over $\uf$'s obeying 
$G(\uf , \uf) = \bfone$ can be relaxed to the quadratic
form inequality $0 \leq G(\uf , \uf) \leq \bfone$ without
changing the minimum. This observation foreshadows Lieb's
variational principle \cite{Lieb1981a} formulated shortly after. 
As opposed to the set of $\uf$'s obeying $G(\uf , \uf) = \bfone$, the
set of $\uf$'s obeying the weaker constraint $0 \leq G(\uf , \uf) \leq
\bfone$ is weakly closed (in the appropriate topology) which is 
necessary for the application of weak lower semicontinuity.

The HVZ-type condition $E_\HF(N) < E_\HF(N-1)$ mentioned above is the
key condition for the proof of existence of excited states of Coulomb
systems in Hartree--Fock theory, too. As the latter leads to nonlinear
Euler-Lagrange equations, the concept of excited state as a higher
eigenvalue of a linear operator cannot be applied directly, but there
is a natural notion for \textit{excited states} in variational
analysis, namely, stationary points of the functional under
consideration for values strictly above the minimum. The first proof
that such excited states exist was given by Lions in \cite{Lions1987}.
\BBBLUE{Building} up on a contribution by Friesecke
\cite{Friesecke2003} and an earlier paper \cite{Lewin2011}, Lewin
proved in \cite{Lewin2018} the existence of infinitely many excited
states below $E_\HF(N-1)$. The essential step is to prove that below
$E_\HF(N-1)$, the the Hartree--Fock functional for Coulomb systems
fulfills a suitable Palais--Smale condition. \BBBLUE{Moreover,
  \RED{unlike} \cite{LiebSimon1977b} and \cite{Lions1987}, the proof
  in \cite{Lewin2011, Lewin2018} is entirely given in the space $N$
  electrons and does not use any positivity of the pair potential or
  its Fourier transform.}

\newpage
\section{Fock Space, Density Matrices, \\ 
and Second Quantization} 
\label{sec-II}
%
Before we turn
to Lieb's variational principle we provide a convenient
mathematical framework and introduce the second quantization.

\paragraph*{Fock Space:}
We henceforth assume the one-particle Hilbert space $\fh$ 
to be a complex separable Hilbert space - not necessarily 
$L^2\big( \RR^3 \times \{\uparrow, \downarrow\} \big)$ as specified
in \eqref{eq-I.05}, although this is a good example to keep in mind.
For $N \in \ZZ^+$, the \RED{\textit{$\RED{N}$-particle Hilbert space}}
is
\begin{align} \label{eq-II.01}
\RED{\fH^{(N)}} 
\ := \ & 
\bigwedge^N \fh 
\ := \ 
\ol{ \huelle\Big\{ f_1 \wedge \cdots \wedge f_N \: \Big| \ 
f_1, \ldots, f_N \: \in \: \fh \Big\} }^{\, \| \cdot \|}
\ \subseteq \ 
\fh^{\otimes N} \, ,
\end{align}
where $\ol{ ( \: \cdot \: ) }^{\, \| \cdot \|}$ denotes norm closure.
The \RED{\textit{fermion Fock space}} (over $\fh$) is defined to be the
orthogonal sum
\begin{align} \label{eq-II.02}
\RED{\fF} \ \equiv \ \RED{\fF[\fh]}
\ := \ & 
\bigoplus_{N=0}^\infty \fH^{(N)} \, ,
\end{align}
where $\RED{\fH^{(0)}} := \CC \cdot \Om$ is the one-dimensional
\RED{\textit{vacuum subspace}} spanned by a unit vector
$\RED{\Om}$ called the \RED{\textit{vacuum vector}}. The vacuum
subspace represents the physical state of absence of any particle in
the quantum system under consideration. The elements of $\fF$ are
sequences $\Psi = (\psi_0, \psi_1, \psi_2, \ldots)$ with 
$\psi_N \in \fH^{(N)}$. If no confusion is possible, we henceforth
consider $\fH^{(N)}$ a subspace of $\fF$ by identifying 
$\psi_N \in \fH^{(N)}$ with 
$(0, \ldots, 0, \psi_N, 0, \ldots) \in \fF$.

\paragraph*{Second Quantization:}
We come to the second quantization of operators. Given 
$N \geq 2$ and three indices $i,j, k \in \{1, \ldots, N\}$, $i < j$, 
we define two unitary operators $\Pi_{i}^{(N)} \in \cU(\fh^{\otimes N})$
and $\Pi_{i,j}^{(N)} \in \cU(\fh^{\otimes N})$ by 
\begin{align} \label{eq-II.03}
\Pi_{k}^{(N)} & [ f_1 \otimes \cdots \otimes f_N ]
\ := \ 
f_k \otimes f_1 \otimes \cdots \otimes 
f_{k-1} \otimes f_{k+1} \otimes \cdots \otimes f_N \, ,
\\[1ex] \label{eq-II.04}
\Pi_{i,j}^{(N)} & [ f_1 \otimes \cdots \otimes f_N ]
\ := \
\\ \nonumber 
& f_i \otimes f_j \otimes f_1  \otimes \cdots \otimes 
f_{i-1} \otimes f_{i+1} \otimes \cdots \otimes 
f_{j-1} \otimes f_{j+1} \otimes \cdots \otimes f_N \, .
\end{align}
Next, given a one-particle operator $h$ on $\fh$ and a two-particle
operator $V$ on \BLUE{$\fh \otimes \fh$}, we define the corresponding
$N$-particle operators $h_N$, $V_N$ and furthermore $H_N$ 
on $\fH^{(N)}$ by $h_0 := V_0 := 0$, $h_1 := h$, $V_1 := 0$, and
\begin{align} \label{eq-II.05}
h_N \ := \ &
\sum_{i=1}^N \big( \Pi_{i}^{(N)} \big)^* \,
\big( h \otimes \bfone^{\otimes (N-1)} \big) \, \Pi_{i}^{(N)}  
\\[1ex] \label{eq-II.06}
V_N \ := \ &
2 \sum_{1 \leq i < j \leq N} \big( \Pi_{i,j}^{(N)} \big)^* \,
\big( V \otimes \bfone^{\otimes (N-2)} \big) \, \Pi_{i,j}^{(N)} \, ,
\\[1ex] \label{eq-II.07}
H_N \ := \ & h_N \: + \: \tfrac{1}{2} V_N \, .  
\end{align}
Note that $H_N$ agrees with the operator in \eqref{eq-I.02}
provided that $h = -\Delta_x - \sum_{k=1}^K Z_k |\vx - \vR_k|^{-1}$
and $V = |\vx - \vy|^{-1}$.
Their \RED{\textit{second quantizations}} are the operators 
\begin{align} \label{eq-II.08}
\RED{\hh} \ := \ \bigoplus_{N=0}^\infty h_N \, , \quad
\RED{\VV} \ := \ \bigoplus_{N=0}^\infty V_N \, , \quad
\RED{\HH} \ := \ \bigoplus_{N=0}^\infty H_N 
\ = \ \hh + \tfrac{1}{2} \VV \, ,
\end{align}
acting on finite vectors [defined in \eqref{eq-II.16}]. The
question whether $\HH$ extends to a semibounded qua\-dra\-tic form is
subtle, in general. For Coulomb systems, however, Dyson and Lenard
\cite{DysonLenard1967a, DysonLenard1967b} and Lieb and Thirring
\cite{LiebThirring1975} have shown \textit{stability of matter} to
hold true, which in our context means that
\begin{align} \label{eq-II.08,01}
\HH_\mu \ := \ 
\HH \: - \: \mu \, \NN
\ \geq \ 
\mu \, Z \: - \: 
\sum_{1 \leq k < \ell \leq K} Z_k \, Z_\ell \, |R_k-R_\ell|^{-1} \, ,
\end{align}
as a quadratic form on $\fF$, provided that the 
\RED{\textit{chemical potential}} $\RED{\mu} < 0$ is
chosen sufficiently small, where
\begin{align} \label{eq-II.09}
\RED{\NN} \ := \ \bigoplus_{N=0}^\infty N \cdot \bfone_{\fH^{(N)}} 
\end{align}
is the \RED{\textit{number operator}} on $\fF$. \BBBLUE{(See also 
\RED{Loss'} contribution to this volume.)}

Abstracting from this situation, \RED{in the following we assume} the
operator $h$ to be essentially self-adjoint on a suitable
dense domain $\fs \subseteq \fh$ and
semibounded, so that $h(m) := h + m \geq 0$, for some sufficiently
large constant $m \in \RR$. Furthermore, the pair potential is assumed
to be an infinitesimal perturbation of $h$, i.e., $V$ is defined on
$\fs$ and, for any $\eps >0$, there exists a constant 
$b_\eps < \infty$, such that 
$\|V f \|_\fh \leq \eps \|h f \|_\fh + b_\eps \|f\|_\fh$ holds true
for all $f \in \fs$.

\paragraph*{Density Matrices:}
The energy expectation value $\la \psi_N \, | \, H_N \psi_N \ra$ of a
state represented by an $N$-particle wave function 
$\psi_N \in \fH^{(N)}$ may be written as 
$\la \Psi \, | \, \HH \Psi \ra$, where 
$\Psi = (0, \ldots, 0, \psi_N, 0, \ldots ) \in \fF$ has only one
non-vanishing component. Allowing for linear combinations,
it can be extended to all finite 
vectors $\Psi \in \fF_\fin$ of sufficient regularity. For these, we
can further rewrite $\la \Psi \, | \, \HH \Psi \ra = \Tr_\fF\big(
\HH \: |\Psi \ra\la \Psi| \big)$, where 
$|\Psi \ra\la \Psi| \in \cB(\fF)$ denotes the rank-one orthogonal
projection onto $\Psi$. 

This suggests \RED{we} further extend the notion of energy expectation value
to all \RED{\textit{density matrices}}
\begin{align} \label{eq-II.10}
\RED{\DM} 
\ := \ 
\Big\{ \rho \in \cL^1(\fF) \; \Big| 
\ \rho \geq 0 \, , \ \ \Tr_\fF(\rho) \: = \: 1 \, , \ \  
\text{$\rho$ is even} \Big\} \, ,
\end{align}
i.e., all even positive trace-class operators $\rho$ on $\fF$ of unit
trace. \BBBLUE{(Here and henceforth we use the convention that $a \geq
  0$ includes the self-adjointness of an operator $a$.)} Evenness of
$\rho$ means that $\la \phi | \rho \psi \ra = 0$, whenever $\phi \in
\fH^{(m)}$ and $\psi \in \fH^{(n)}$ with $m-n$ odd. We remark that
evenness of density matrices is a natural condition for fermion, but
not for boson systems.
Since a given density matrix $\rho \in \DM$ is, in particular,
self-adjoint and compact, it can be written in diagonal form as 
$\rho = \sum_{\nu=1}^\infty r_\nu \, |\Psi_\nu \ra\la \Psi_\nu|$,
where $r_\nu \geq 0$ are its nonnegative eigenvalues, which \RED{sum to $1$}, 
and $\{ \Psi_\nu \}_{\nu=1}^\infty \subseteq \fF$ is an
orthonormal basis in $\fF$ of eigenvectors of $\rho$. If a density
matrix $\rho \in \DM$ obeys $\rho \NN = \NN \rho = N \cdot \rho$, for
some $N \in \ZZ^+$, then $\rho$ is called an
\RED{\textit{$\RED{N}$-particle density matrix}}. These are collected
in
\begin{align} \label{eq-II.11}
\RED{\DM^{(N)}} 
\ := \ 
\big\{ \rho \in \DM \; \big| 
\ \rho \NN = \NN \rho = N \cdot \rho \big\} \, ,
\end{align}
Note that $\rho \in \DM^{(N)}$ if, and only if, all its eigenvectors
belong to $\fH^{(N)}$. Moreover, the density matrices form a
norm-closed convex subset $\DM \subseteq \cL^1(\fF)$ in which rank-one
orthogonal projections, such as $|\Psi \ra\la \Psi| \in \DM$ above,
are extremal points called \RED{\textit{pure}}. In particular, the
orthogonal projection onto $\psi_N \in \fH^{(N)}$ considered above 
is a pure $N$-particle density matrix.

Density matrices (and, in general, states) are not only natural objects
mathematically, but from a physics point of view they are also
important conceptually: For any reasonable theoretical framework for
the description of a physical system, the scenario that this system is
a subsystem of a larger system (``the rest of the universe'') ought to
be built in. In the latter situation, however, density matrices
resulting from projecting onto the subsystem are the natural physical
states, not wave functions.

Equipped with the definitions of density matrices and $N$-particle
density matrices above, the Rayleigh--Ritz principle \eqref{eq-I.03}
assumes the following abstract form:
\begin{align} \label{eq-II.12}
E_\gs(N) 
\ = \  
\inf\big\{ \Tr_\fF( \rho \, \HH ) \; \big| 
\ \rho \: \in \: \DM^{(N)} \, , \ \la \HH \ra_\rho < \infty \big\} \, ,
\end{align}
where it is implicity assumed that $H_N$ is bounded from below and we
denote 
\begin{align} \label{eq-II.12,01}
\la A \ra_\rho 
\ := \ 
\Tr_\fF( \rho^{1/2} \, A \, \rho^{1/2}) \, .
\end{align}
Similarly, for a sufficiently small chemical potential $\mu < 0$ such
that $\HH_\mu$ is bounded below, we define the \RED{\textit{total ground
  state energy}}
\begin{align} \label{eq-II.12,02}
E_\gs 
\ := \  
\inf_{N \in \ZZ^+} E_\gs(N) 
\ = \  
\inf\big\{ \Tr_\fF( \rho \, \HH_\mu ) \; \big| 
\ \rho \: \in \: \DM \, , \ \la \HH \ra_\rho < \infty \big\} \, .
\end{align}

\paragraph*{Creation and Annihilation Operators:}
Next, we introduce creation operators. Fixing $f \in \fh$, we define
$c^*(f): \fH^{(0)} \to \fH^{(1)}$ by $c^*(f)\Om := f$ and 
$c^*(f): \fH^{(N)} \to \fH^{(N+1)}$, for $N \in \ZZ^+$, by
\begin{align} \label{eq-II.13}
c^*(f) [ f_1 \wedge \cdots \wedge f_N ] 
\ := \ 
f \wedge f_1 \wedge \cdots \wedge f_N
\end{align}
and extension by linearity. One then easily checks that $c^*(f)$
extends by continuity to a bounded operator on $\fF$ of norm 
$\| c^*(f) \|_{\cB(\fF)} = \|f\|_\fh$, \GREEN{called}
\RED{\textit{creation operator}}
$\RED{c^*(f)} \in \cB(\fF)$. We observe that
\begin{align} \label{eq-II.14}
\fH^{(N)}
\ = \ &
\ol{ \huelle\Big\{ c^*(f_1) \cdots c^*(f_N) \Om \: \Big| \ 
f_1, \ldots, f_N \: \in \: \fh \Big\} }^{\, \| \cdot \|}
\quad \text{and} 
\\[1ex] \label{eq-II.15}
\fF
\ = \ &
\ol{ \huelle\Big\{ c^*(f_1) \cdots c^*(f_N) \Om \: 
\Big| \ N \in \ZZ_0^+ \, ,
\ \ f_1, \ldots, f_N \: \in \: \fh \Big\} }^{\, \| \cdot \|} \, ,
\end{align}
for $N \in \ZZ_0^+ := \{0,1,2,3,\ldots\}$. Note that if 
$\fs \subseteq \fh$ is a dense subspace then the \RED{\textit{space}}
\begin{align} \label{eq-II.16}
\RED{\fF_\fin[\fs]}
\ := \ &
\huelle\Big\{ c^*(f_1) \cdots c^*(f_N) \Om \: 
\Big| \ N \in \ZZ_0^+ \, ,
\ \ f_1, \ldots, f_N \: \in \: \fs \Big\} 
\ \subseteq \ \fF[\fh]
\end{align}
\RED{\textit{of finite vectors}} in $\fF$ containing finite linear
combinations of finite wedge-products of orbitals in $\fs$ 
is a convenient dense domain for second quantizations of
one- and two-particle operators on $\fF$.

The adjoint $\RED{c(f)} := [ c^*(f) ]^* \in \cB(\fF)$ of 
$c^*(f)$ is called \RED{\textit{annihilation operator}}. We remark that, 
\RED{although} $\fh \ni f \mapsto c^*(f) \in \cB(\fF)$ is linear, 
the map $\fh \ni f \mapsto c(f) \in \cB(\fF)$ is antilinear.
Moreover, \RED{it can be easily checked} that the family 
$\{ c^*(f), c(f) \}_{f \in \fh} \subseteq \cB(\fF)$ 
of creation and annihilation operators define a
\RED{\textit{Fock representation}} of the 
\RED{\textit{canonical anticommutation relations (CAR)}}, i.e., 
it fulfills 
\begin{align} \label{eq-II.17}
\big\{ c(f) , c^*(g) \big\} 
\ = \
\la f | g \ra_\fh \cdot \bfone_{\fF} 
\quad \text{and} \quad c(f) \Om = 0 \, , 
\end{align}
for all $f, g \in \fh$, where $\{ a, b \} := ab + ba$ is the anticommutator
of two operators $a$ and $b$.

The second-quantized operators $\hh$, $\VV$, $\HH$, and $\NN$ have a 
convenient representation in terms of creation and annihilation 
operators. Assuming that $\fh$ is infinite-dimensional and
that $\{f_j\}_{j=1}^\infty \subseteq \fs \subseteq \fh$ is an
orthonormal basis of $\fh$ of sufficiently regular functions lying in 
a dense subspace $\fs$, we have that
\begin{align} \label{eq-II.18}
\NN \ = \ 
\sum_{j=1}^\infty c^*(f_j) \, c(f_j) \, , & \quad
\hh \ = \ 
\sum_{j,k=1}^\infty \la f_j | \, h f_k \ra \, c^*(f_j) \, c(f_k) \, ,
\\[1ex] \label{eq-II.19}
\VV \ = \
\sum_{i,j,k,\ell =1}^\infty 
\la f_i  \otimes f_j | \, & V ( f_k \otimes f_\ell) \ra 
\, c^*(f_i) \, c^*(f_j) \, c(f_\ell) \, c(f_k) \, ,
\end{align}

\paragraph*{Reduced Density Matrices:}
We come to defining reduced density matrices - the central object of
this paper. Given a density matrix $\rho \in \DM$ with finite
expectation value $\la \NN \ra_\rho < \infty$ of $\NN$, we
define its \RED{\textit{reduced one-particle density matrix (1-RDM)}}
$\RED{\gamma_\rho^{(1)}}$ as a linear operator on $\fh$. 
Likewise, if $\la \NN^2 \ra_\rho < \infty$ we define 
its \RED{\textit{reduced two-particle density matrix (2-RDM)}} 
$\RED{\gamma_\rho^{(2)}}$ as a linear operator on 
$\fh \otimes \fh$ by their matrix elements
\begin{align} \label{eq-II.20}
\big\la g \, \big| \; \gamma_\rho^{(1)} f \big\ra_\fh
\ := \ &
\Tr_\fF\big[ \rho \, c^*(f) \, c(g) \big] \, , 
\\[1ex] \label{eq-II.21}
\big\la g_1 \otimes g_2 \, \big| \; 
\gamma_\rho^{(2)} (f_1 \otimes f_2) \big\ra_{\fh \otimes \fh}
\ := \ &
\Tr_\fF\big[ \rho \, c^*(f_1) \, c^*(f_2) \, c(g_2) \, c(g_2) \big] \, ,
\end{align}
for all $f, g, f_1, f_2, g_1, g_2 \in \fh$. These definitions are
meaningful because it turns out that they define trace-class
operators, as the following lemma asserts.
%
\begin{lemma} \label{lem-II.01}
Let $\rho \in \DM$ be a density matrix of 
finite particle number variance 
$\la \NN^2 \ra_\rho := \Tr_\fF[\rho \, \NN^2 ] < \infty$, and define its
1-RDM $\gamma_\rho^{(1)}$ and its 2-RDM 
$\gamma_\rho^{(2)}$ by \eqref{eq-II.20} and \eqref{eq-II.21}, 
respectively. Then $\gamma_\rho^{(1)}$ and
$\gamma_\rho^{(2)}$ possess the following properties:
\begin{itemize}
\item[(i)] 
The operators $\gamma_\rho^{(1)} \in \cL^1(\fh)$ and 
$\gamma_\rho^{(2)} \in \cL^1(\fh \otimes \fh)$ 
are positive trace-class operators of trace
\begin{align} \label{eq-II.22}
\Tr_{\fh} [ \gamma_\rho^{(1)} ] 
\ = \ 
\big\la \NN \big\ra_\rho  
\quad \text{and} \quad
\Tr_{\fh \otimes \fh} [ \gamma_\rho^{(2)} ] 
\ = \ 
\big\la \NN^2 - \NN \big\ra_\rho \, .
\end{align}

\item[(ii)] 
As quadratic forms, 
\begin{align} \label{eq-II.23}
0 \ \leq \ \gamma_\rho^{(1)} \ \leq \ \bfone_\fh
\quad \text{and} \quad
0 \ \leq \ \gamma_\rho^{(2)} \ \leq \ (N-1) \, \bfone_{\fh \otimes \fh} \, .
\end{align}

\item[(iii)] 
Suppose that $h$ and $V$ are semibounded and 
$\la \hh \ra_\rho, \la \VV \ra_\rho < \infty$. Then 
\begin{align} \label{eq-II.24}
\big\la \HH \big\ra_\rho 
\ = \ 
\cE_Q\big( \gamma_\rho^{(1)}, \gamma_\rho^{(2)} \big)
\ := \
\Tr_\fh\big[ h \, \gamma_\rho^{(1)} \big] \: + \: \tfrac{1}{2}
\Tr_{\fh \otimes \fh}\big[ V \, \gamma_\rho^{(2)} \big] \, .
\end{align}
\end{itemize}
\end{lemma}

We finally remark that, in case of an $N$-particle density matrix
$\rho \in \DM^{(N)}$, for some $N \geq 2$, the corresponding 1-RDM
$\gamma_\rho^{(1)}$ can be obtained from its 2-RDM $\gamma_\rho^{(2)}$
by taking a partial trace,
\begin{align} \label{eq-II.25}
\big\la g \, \big| \; \gamma_\rho^{(1)} f \big\ra_\fh
\ = \ &
\frac{1}{N-1} \, \sum_{j=1}^\infty 
\big\la g \otimes f_j \, \big| \; 
\gamma_\rho^{(2)} (f \otimes f_j) \big\ra_{\fh \otimes \fh} \, ,
\end{align}
where $\{ f_j \}_{j=1}^\infty \subseteq \fh$ is an orthonormal basis.

\newpage
\section{Lieb's Variational Principle} 
\label{sec-III}
%
The formulation of the Hartree--Fock approximation in
Section~\ref{sec-I} turns out to be too rigid and inconvenient,
mathematically. A more flexible formulation is provided by Lieb's
variational principle which uses the reduced one-particle
density matrices introduced in \eqref{eq-II.20}. 

We first link the 1-RDM $\gamma_\rho^{(1)}$ corresponding to a density
matrix $\rho \in \DM$ to the one-particle density matrix
$\gamma_{\uf}$ corresponding to an orthonormal family 
$\uf = (f_1, \ldots, f_N) \in \fh^N$ of $N$ orbitals that enter the
Hartree--Fock energy functional $\cE_\HF(\uf)$ in \eqref{eq-I.07}. In
fact, for pure states we have the following important relation between
these operators:
%
\begin{lemma} \label{lem-III.01}
For $N \geq 2$, let $\uf = (f_1, \ldots, f_N) \in \fh^N$ 
with $G(\uf, \uf) = \bfone$ and assume that 
$\rho(\uf) = |\Phi(\uf) \ra\la \Phi(\uf)| \in \DM^{(N)}$ is the
orthogonal projection onto the Slater determinant 
$\Phi(\uf) = f_1 \wedge \cdots \wedge f_N$. 
Then the following statements hold true.
\begin{itemize}
\item[(i)] The reduced one-particle density matrix of $\rho(\uf)$ is the 
rank-$N$ orthogonal projection
\begin{align} \label{eq-III.01}
\gamma_{\rho(\uf)}^{(1)} \ = \ \sum_{i=1}^N |f_i\ra\la f_i| 
\end{align}
onto the subspace of $\fh$ spanned by $\{ f_1, \ldots, f_N \}$.

\item[(ii)] The reduced two-particle density matrix of $\rho(\uf)$
is \BBBLUE{twice} the orthogonal projection of rank $\frac{1}{2}N(N-1)$ 
\BBBLUE{\sssout{times two}}, 
\begin{align} \label{eq-III.02}
\gamma_{\rho(\uf)}^{(2)} 
\ = \ 
\sum_{i,j=1}^N |f_i \wedge f_j \ra\la f_i \wedge f_j| 
\ = \ 
\big( \bfone - \Ex \big) 
\big( \gamma_{\rho(\uf)}^{(1)} \otimes \gamma_{\rho(\uf)}^{(1)} \big) \, ,
\end{align}
onto the subspace of $\fh \otimes \fh$ spanned by 
$\{ f_i \wedge f_j | 1 \leq i < j \leq N \}$. Here, 
$\RED{\Ex} \in \cU(\fh \otimes \fh)$ denotes the 
\RED{\textit{exchange operator}} $f \otimes g \mapsto g \otimes f$ 
\end{itemize}
\end{lemma}
%
Inserting \eqref{eq-III.02} into \eqref{eq-II.24}, we immediately
obtain
\begin{align} \label{eq-III.10}
\la \Phi(\uf) \, | \; H_N \Phi(\uf) \ra 
\ = \ &
\cE_Q\big( \gamma_{\rho(\uf)}^{(1)}, \; \BLUE{\text{insert 1mm space}} 
(\bfone - \Ex)(\gamma_{\rho(\uf)}^{(1)} \otimes \gamma_{\rho(\uf)}^{(1)}) \big)
\\[1ex] \nonumber
\ = \ & 
\Tr_\fh\big[ h \, \gamma_{\rho(\uf)}^{(1)} \big] \: + \:
\frac{1}{2} \, \Tr_{\fh \otimes \fh} \big[ V \, (\bfone - \Ex)
(\gamma_{\rho(\uf)}^{(1)} \otimes \gamma_{\rho(\uf)}^{(1)}) \big] \, ,
\end{align}
if the energy expectation is evaluated on a Slater determinant
$\Phi(\uf) = f_1 \wedge \cdots \wedge f_N$. The right side of
\eqref{eq-III.10} is entirely expessed in terms of the rank-$N$
orthogonal projection $\gamma_{\rho(\uf)}^{(1)}$, and no other
property than that enters the functional. That is, 
\begin{align} \label{eq-III.11}
E_\HF(N) 
\ = \  
\inf\Big\{ \cE_\HF( \gamma ) \ \Big| \ 
\gamma = \gamma^* \in \cL^1(\fh) \, , \ 
\Tr(\gamma) = N \, , \ \gamma = \gamma^2 \Big\} \, ,
\end{align}
where 
\begin{align} \label{eq-III.12}
\cE_\HF( \gamma ) 
\ = \ 
\Tr_\fh[ h \, \gamma ] \: + \:
\frac{1}{2} \, \Tr_{\fh \otimes \fh} [ V \, (\bfone - \Ex)
(\gamma \otimes \gamma) ] \, .
\end{align}
Lieb's variational principle \cite{Lieb1981a} asserts that the
projection property $\gamma = \gamma^2$ in \eqref{eq-III.11} can be
relaxed to $0 \leq \gamma \leq \bfone_\fh$ without changing the
infimum of the functional.
%
\begin{theorem}[Lieb's Variational Principle] \label{thm-III.03}
For $N \geq 2$, 
\begin{align} \label{eq-III.13}
E_\HF(N) 
\ = \  
\inf\Big\{ \cE_\HF( \gamma ) \ \Big| \ 
\gamma \ \in \cL^1(\fh) \, , \ 
\Tr(\gamma) = N \, , \ 0 \leq \gamma \leq \bfone_\fh \Big\} \, .
\end{align}
\begin{proof} We define the auxiliary
energy $E_\aux(N)$ to be the infimium on the right side in
\eqref{eq-III.13} and observe that, clearly, $E_\aux(N) \leq E_\HF(N)$. 
We make three simplifying assumptions which are not
essential for the validity of Theorem~\ref{thm-III.03}
and can be avoided by suitable limiting arguments. The first is
the strict positivity of $\la \psi | V \psi \ra > 0$, for nonvanishing
$\psi$, as opposed to merely assuming
$\la \psi | V \psi \ra \geq 0$. The second simplifying assumption 
we make is that the infimum $E_\aux(N)$ 
is actually a minimum. That is, $E_\aux(N) = \cE_\HF(\gamma_0)$ is 
attained by a minimizer $\gamma_0$ which fulfills 
$0 \leq \gamma_0 \leq \bfone$ and $\Tr(\gamma_0) = N$. Since
$\gamma_0$ is compact, there exist an orthonormal basis
$\{f_i\}_{i=1}^\infty \subseteq \fh$ of eigenvectors of $\gamma_0$
with corresponding (not necessarily distinct) eigenvalues 
$\lambda_i \in [0,1]$ that sum up to $N$. The third assumption
we make is that $\gamma_0$ is of finite rank $J < \infty$, so that 
$\lambda_1, \ldots, \lambda_J >0$,
\begin{align} \label{eq-III.14}
\gamma_0 \ = \ \sum_{j=1}^J \lambda_j | f_j \ra\la f_j |
\quad \text{and} \quad
\cE_\HF(\gamma_0) \ = \ 
\sum_{i=1}^J \lambda_i \, h_i \: + \: 
\frac{1}{2} \sum_{i,j=1}^J \lambda_i \, \lambda_j \, V_{i,j} \, ,
\end{align}
where $h_i := \la f_i | h f_i \ra$ and 
$V_{i,j} := \la f_i \wedge f_j | V (f_i \wedge f_j) \ra > 0$.

Before we turn to Lieb's original proof in \cite{Lieb1981a} we sketch
the proof that $E_\aux(N) \geq E_\HF(N)$ given in
\cite{Bach1992}\BBBLUE{, which, however, takes a different perspective}.
First note that it suffices to show that $\gamma_0 = \gamma_0^2$ is a
projection. To this end we assume that $\gamma_0$ is not a projection
and derive a contradiction from this assumption. If $\gamma_0$ is not
a projection then there are (at least) two indices $p, q \in \ZZ^+$,
$p < q$ such that $\lambda_p, \lambda_q \in (0,1)$ because the sum
$\sum_{j=1}^J \lambda_j = N$ is an integer. We set $r :=
\min\{\lambda_p, 1-\lambda_p, \lambda_q, 1-\lambda_q\} >0$ and $I :=
[-r,r]$ and observe that $\lambda_p + \delta, \lambda_q + \delta \in
[0,1]$, for any $\delta \in I$. We define
\begin{align} \label{eq-III.15}
\gamma_\delta \ := \  
(\lambda_p + \delta) | f_p \ra\la f_p | \: + \:
(\lambda_q - \delta) | f_q \ra\la f_q | \: + \:
\sum_{j \in \ZZ^+ \setminus \{p,q\} } \lambda_j | f_j \ra\la f_j | \, .
\end{align}
Then $0 \leq \gamma_\delta \leq \bfone$ and $\Tr(\gamma_\delta) = N$,
so $\gamma_\delta$ is admissible for any $\delta \in I$. Moreover
$\gamma_0 = \frac{1}{2} \gamma_\delta + \frac{1}{2} \gamma_{-\delta}$.
A simple computation using that $V_{p,q} >0$ shows the strict concavity of 
$I \ni \delta \mapsto \cE_\HF(\gamma_\delta)$. Hence, 
$\min\{\cE_\HF(\gamma_r),\cE_\HF(\gamma_{-r})\} < \cE_\HF(\gamma_0)$,
which contradicts the assumption that $\gamma_0$ is a minimizer of
$\cE_\HF$. It follows that $\gamma_0$ is a projection, indeed. 
Note that this proof is constructive in the sense that, 
fixing the orthonormal orbitals $f_1, \ldots, f_J$, 
it defines an algorithm to find the $\gamma_*$ of minimal energy
$\cE_\HF(\gamma_*)$ among all rank-$J$ operators of the form 
$\gamma(\tau_1, \ldots, \tau_J) \ = \ \sum_{j=1}^J \tau_j | f_j \ra\la f_j |$
with $0 \leq \tau_j \leq 1$ and $\sum_{j=1}^J \tau_j = N$.

We now turn to Lieb's proof of Theorem~\ref{thm-III.03} in
\cite{Lieb1981a}, starting from \eqref{eq-III.14}. Its heart
is a lemma that, under the assumption that 
$\lambda_1, \ldots, \lambda_J >0$ and 
$\lambda_1 + \ldots + \lambda_J =N$,
asserts the existence of $N$ orthonormal vectors 
$G^{(1)}, \ldots, G^{(N)} \in \CC^J$ which fulfill 
$\sum_{n=1}^N |G_j^{(n)}|^2 = \lambda_j$, for all 
$j \in \{1, \ldots, J\}$. We omit its interesting proof.
Given these vectors $G^{(1)}, \ldots, G^{(N)}$, Lieb defines 
\begin{align} \label{eq-III.16}
g_n^{(\theta)} \ := \ \sum_{j=1}^J e^{2\pi i \theta_j} \: G_j^{(n)} \: f_j
\ \in \ \fh \, ,
\end{align}
for all $n \in \{1, \ldots, N\}$ and any choice 
$\theta := (\theta_j)_{j=1}^J \in [0,1)^J$ of phases 
$\theta_1, \ldots, \theta_J$. Using the orthonormality of 
$\{f_1, \ldots, f_J \} \subseteq \fh$
it is easy to check that the set
$\{ g_1^{(\theta)}, \ldots, g_N^{(\theta)} \} \subseteq \fh$
is orthonormal, too. That is, $G(\ug^{(\theta)}, \ug^{(\theta)}) = \bfone$,
where $\ug^{(\theta)} = (g_1^{(\theta)} \ldots, g_N^{(\theta)}) \in \fh^N$,
and the corresponding Slater determinant is
$\Phi(\ug^{(\theta)}) = g_1^{(\theta)} \wedge \cdots \wedge
g_N^{(\theta)} \in \fH^{(N)}$. The energy expectation value of this 
Slater determinant is 
\begin{align} \label{eq-III.17}
\big\la \Phi(\ug^{(\theta)}) \, & \big| \, H_N \Phi(\ug^{(\theta)}) \big\ra 
\\[1ex] \nonumber 
\ = \ &
\sum_{n=1}^N \la g_n^{(\theta)} \, | \, h g_n^{(\theta)} \ra_\fh 
\; + \; \frac{1}{2} \sum_{m,n=1}^N 
\la g_m^{(\theta)} \wedge g_n^{(\theta)} \, | \, 
V (g_m^{(\theta)} \wedge g_n^{(\theta)}) \ra_{\fh \otimes \fh} \, .
\end{align}
This energy expectation value is now averaged over all possible choices 
of $\theta$ by integrating over $[0,1)^J$. That is, for any integrable
function $F \in L^1( [0,1)^J )$ we write 
$\EE_\theta[F] := \int_0^1 \cdots \int_0^1 
F(\theta) \: d\theta_1 \cdots d\theta_J$. Using that 
$\EE_\theta[ e^{2\pi i (\Theta_j-\Theta_k)} ] = \delta_{j,k}$ and \eqref{eq-III.16}, 
Lieb obtains
\begin{align} \label{eq-III.18}
\EE_\theta\bigg[ 
\sum_{n=1}^N \la g_n^{(\Theta)} \, | & \, h g_n^{(\Theta)} \ra_\fh \bigg]
\ = \ 
\sum_{n=1}^N \sum_{j,k=1}^J \EE_\theta\big[ e^{2\pi i (\Theta_j-\Theta_k)}\big] 
\: G_j^{(n)} \, \ol{G_k^{(n)}} \: \la f_j \, | \, h f_k \ra_\fh
\nonumber \\[1ex] 
\ = \ &
\sum_{j=1}^J \bigg( \sum_{n=1}^N |G_j^{(n)}|^2 \bigg) 
\: \la f_j \, | \, h f_j \ra_\fh
\ = \
\sum_{j=1}^J \lambda_j \: h_j \, .
\end{align}
Similarly, if $i \neq j$ and $k \neq \ell$ then 
$\EE_\theta[ e^{2\pi i (\Theta_i+\Theta_j-\Theta_k-\Theta_\ell)} ] = 
(\delta_{i,k} \, \delta_{j,\ell} + \delta_{i,\ell} \, \delta_{j,k})$,
and this implies that
\begin{align} \label{eq-III.19}
\EE_\theta\bigg[
\sum_{m,n=1}^N & 
\la g_m^{(\Theta)} \wedge g_n^{(\Theta)} \, | \, 
V (g_m^{(\Theta)} \wedge g_n^{(\Theta)}) \ra_{\fh \otimes \fh} \bigg]
\\[1ex] \nonumber 
\ = \ &
\sum_{m,n=1}^N \sum_{i,j,k,\ell =1}^J 
\EE_\theta\big[ e^{2\pi i (\Theta_i+\Theta_j-\Theta_k-\Theta_\ell)}\big] 
\\ \nonumber 
& \qquad G_i^{(m)} \, G_j^{(n)} \,  \ol{G_k^{(m)} \, G_\ell^{(n)}} \: 
\la f_i \wedge f_j \, | \, V ( f_k \wedge f_\ell) \ra_{\fh \otimes \fh}
\\[1ex] \nonumber 
\ = \ &
\sum_{m,n=1}^N \sum_{i,j =1}^J \Big( |G_i^{(m)}|^2 \, |G_j^{(n)}|^2
\: - \: G_j^{(m)} \, G_i^{(n)} \,  \ol{G_i^{(m)} \, G_j^{(n)}} \Big) \, V_{i,j}
\\[1ex] \nonumber 
\ = \ &
\sum_{i,j =1}^J \lambda_i \, \lambda_j \, V_{i,j}
\: - \: 
\sum_{i,j =1}^J \bigg| \sum_{n=1}^N G_i^{(n)} \, \ol{G_j^{(n)}} \bigg|^2 
\, V_{i,j} 
\ \leq \ 
\sum_{i,j =1}^J \lambda_i \, \lambda_j \, V_{i,j} \, .
\end{align}
\BBBLUE{Here, the positivity $V \geq 0$ is crucial, see also
  \eqref{eq-IV.01,1}.} Adding up \eqref{eq-III.18} and half of
\eqref{eq-III.19}, Lieb arrives at
\begin{align} \label{eq-III.20}
\EE_\theta\big[ \la \Phi(\ug^{(\Theta)}) \, | 
\, H_N \Phi(\ug^{(\Theta)}) \ra \big] 
\ \leq \ 
\cE_\aux(\gamma_0) \, .
\end{align}
Since $\EE_\theta$ is an average, Eq.~\eqref{eq-III.20} implies that
there is at least one choice 
of $\theta \in [0,1)^J$, for which 
$\la \Phi(\ug^{(\theta)}) \big| \, H_N \Phi(\ug^{(\theta)}) \ra 
\leq \cE_\aux(\gamma_0)$. Thus, we finally have $E_\HF(N) \leq E_\aux(N)$.
\end{proof}
\end{theorem}
%
Lieb's variational principle is a formulation of the Hartree--Fock
approximation in the natural variable $\gamma$. It justifies the
introduction of the notion of a \RED{\textit{one-particle density matrix}}
as any self-adjoint trace-class operator 
\begin{align} \label{eq-III.21}
\RED{\gamma} \in \cL^1(\fh)
\quad \text{that obeys} \quad 0 \leq \gamma \leq \bfone_\fh \, , 
\end{align}
leaving aside the question whether it is the reduced one-particle
density matrix $\gamma = \gamma_\rho^{(1)}$ corresponding to some
density matrix $\rho \in \DM$. We come back to this point
in the next section. The one-particle density matrices form
a norm-closed, and hence weakly closed, convex subset of $\cL^1(\fh)$
which makes them suitable for variational analysis.

Lieb's variational principle asserts, briefly speaking, that 
among one-particle density matrices obeying \eqref{eq-III.21}
and of trace $N$ the ones with lowest energy are the rank-$N$
projections. Under the assumption of the existence of a minimizer
$\gamma_\HF$, this conclusion also follows from the stationarity
of the Hartree--Fock functional at $\cE_\HF$ at $\gamma_\HF$.
In fact, the Hartree--Fock equations~\eqref{eq-I.10} turn into
the following \RED{\textit{self-consistent equation}}: 
\begin{align} \label{eq-III.22}
\gamma_\HF \ = \ \bfone_N\big( h_\HF[\gamma_\HF] \big) \, , \ \ 
\BLUE{\text{bigger parentheses}} \end{align}
where $\bfone_N(A)$ denotes the projection onto the lowest $N$
eigenvalues, counting multiplicities, for a self-adjoint operator $A$.
In other words, $\bfone_N(A)$ is the projection onto a subspace of
dimension $N$ such that \BLUE{$\Tr_\fh[ A \cdot \bfone_N(A)]$} is
minimal. (If a minimizer exists, this subspace is actually unique, as
follows from the unfilled-shell theorem of Lieb, Loss, Solovej, and
the author \cite{BachLiebLossSolovej1994}.) Furthermore,
$h_\HF[\gamma]$ is the corresponding form of the Hartree--Fock
effective Hamiltonian, acting on orbitals $g \in \fh$ as
\begin{align} \label{eq-III.23}
\big( h_\HF[\gamma] & \, g \big)[\vx,\tau] \ := \ 
\\[1ex] \nonumber 
& \big( h \, g \big)[\vx,\tau] \: + \:
\bigg( \int \frac{\rho_\gamma(\vy) \, d^3y}{|\vx-\vy|} \bigg) g(\vx,\tau)
\: - \:
\int \frac{\BLUE{\gamma(\vx , \vy)} \, g(\vy,\tau) \: d^3y}{|\vx-\vy|} \, ,
\end{align}
with $\rho_\gamma(\vx) := \gamma(\vx, \vx)$ being the 
one-particle density corresponding to $\gamma$ and a partial trace
$\gamma(\vx , \vy) = \sum_{\tau = \udarrow} 
\gamma(\vx,\tau \, , \, \vy,\tau)$ as well as a sufficiently regular
choice for the integral kernel for $\gamma$ \RED{is} understood. 

Comparing Lieb's variational principle to the original Hartree--Fock
approximation, it is interesting to observe that the
condition $0 \leq G(\uf, \uf) \leq \bfone_{\CC^N}$ considered
by Lieb and Simon in \cite{LiebSimon1977b} is actually 
equivalent to $0 \leq \gamma_\uf \leq \bfone_\fh$, if we set
$\gamma_\uf := \sum_{n=1}^N |f_n\ra\la f_n|$. Note, however, that
$\gamma_\uf$ is of rank $N$, at most, and hence that 
$\Tr_\fh[\gamma_\uf] < N$ unless $\gamma_\uf$ is a rank-$N$ 
projection. It follows that the relaxation of the 
condition $G(\uf, \uf) = \bfone_{\CC^N}$ on the Gram matrix to
the bound $0 \leq G(\uf, \uf) \leq \bfone_{\CC^N}$ is different
from the relaxation of $\gamma = \gamma^2$ to 
$0 \leq \gamma \leq \bfone_\fh$.

\newpage
\section{\GREEN{Bogoliubov} Transformations \\ and Representability}
\label{sec-IV}
%
We begin our discussion of the concept of \textit{representability} by
comparing the two proofs of Theorem~\ref{thm-III.03} given in the
previous section. Lieb's original proof seems to be considerably more
involved than the one in \cite{Bach1992}. One must not overlook,
however, that Lieb proves a stronger statement than
Eq.~\eqref{eq-III.13}. Namely, the averaging procedure introduced
after \eqref{eq-III.17} above yields an $N$-particle density matrix
\begin{align} \label{eq-IV.01}
\rho_\av \ := \ &
\EE_\theta\big( \: |\Phi(\ug^{(\Theta)}) \ra\la \Phi(\ug^{(\Theta)})| \: \big) 
\\[1ex] \nonumber 
\ = \ &
\int_0^1 \cdots \int_0^1 \:  
|\Phi(\ug^{(\theta)}) \ra\la \Phi(\ug^{(\theta)})| 
\: d\theta_1 \cdots d\theta_J 
\ \in \ \DM^{(N)} \, , 
\end{align}
whose reduced one-particle density matrix equals the minimizing
one-particle density matrix $\gamma_0 = \gamma_{\rho_\av}^{(1)}$.
Concerning the energy estimate, the key point in Lieb's construction
is that
\begin{align} \label{eq-IV.01,1}
\gamma_{\rho_\av}^{(2)} \ \leq \ 
( \bfone - \Ex ) \, 
\big( \gamma_{\rho_\av}^{(1)} \otimes \gamma_{\rho_\av}^{(1)} \big) \, ,
\end{align}
which leads to Estimate~\eqref{eq-III.19}, thanks to the positivity 
$V \geq 0$ of the pair interaction potential $V$. To describe the
significance of this observation we introduce some more definitions
and notation. We follow the paper \cite{BachLiebSolovej1994} by Lieb,
Solovej, and the author, Solovej's lecture notes \cite{Solovej2014},
and the papers \cite{BachKnoerrMenge2012, BachBreteauxKnoerrMenge2014,
  BachKnoerrMenge2015, BachHach2021} by Breteaux, Hach, \RED{Knörr, Menge,}
and the author.

\paragraph*{Generalized Reduced Density Matrices} 
The Hamiltonian $\HH$ in \eqref{eq-II.08} is a linear operator on 
$\fF = \bigoplus_{N=0}^\infty \fH^{(N)}$ which leaves the $N$-particle
Hilbert spaces $\fH^{(N)}$ invariant. Thus the variation in 
the Rayleigh-Ritz principles \eqref{eq-II.12,02} for the total ground state
energy $E_\gs$ and \eqref{eq-II.12} for the ground state energy
$E_\gs(N)$ for $N$ particles may both be restricted to density matrices 
$\rho = \bigoplus_{N=0}^\infty \rho_N \in \DM$ that are
particle-number \RED{conserving} and even to
$N$-particle density matrices $\rho \in \DM^{(N)}$ 
without changing the infimum. 

In general, however, density matrices $\rho \in \DM$ need not leave
the $N$-particle Hilbert spaces $\fH^{(N)}$ invariant, they are only
assumed to be even. This can be conveniently formulated with the aid
of the self-dual algebra built from creation and annihilation
operators which was introduced by Araki \cite{Araki1970}. We choose an
antiunitary involution $\sfj: \fh \to \fh$ and define the
\RED{\textit{self-dual field operator}} 
\begin{align} \label{eq-IV.02}
\RED{A^*(f_1 \oplus \sfj f_2 )} \ := \ 
c^*(f_1) + c(f_2) \ \in \ \cB(\fF) \, ,
\end{align}
of a \RED{\textit{generalized orbital $\RED{F} = f_1 \oplus \sfj f_2
  \in \fh \oplus \fh$}}. Neither $A^*(F)$ and $A^*(G)$ nor 
$A(F) := [A^*(F)]^*$ and $A^*(G)$ anticommute, but rather
\begin{align} \label{eq-IV.03}
A(F) \ = \ A^*( \sfJ F )  
\quad \text{and} \quad
\big\{ A(F) , A^*(G) \big\}
\ = \ 
\big\la F \big| G \big\ra_{\fh \oplus \fh} \, ,
\end{align}
where $\sfJ: \fh \oplus \fh \to \fh \oplus \fh$ is the
antiunitary involution defined by 
$\sfJ(f_1 \oplus \sfj f_2) := f_2 \oplus \sfj f_1$. 
All creation and annililation operators can be expessed as 
self-dual field operators $A^*(F)$ for suitable choices 
of $F$. We remark that the antiunitary involution
$\sfj: \fh \to \fh$ ensures the linearity of 
$\fh \oplus \fh \ni F \mapsto A^*(F)$, even
though $\fh \ni f \mapsto c(f)$ is antilinear. 
Its choice is arbitrary and may be adapted to the model
under consideration. The Riesz isomorphism $\fh \to \fh^*$,
$|f\ra \mapsto \la f|$ yields one possible choice. Identifying 
$\fh^*$ with $\fh$, it is the only choice up to unitary transformation
of the domain $\fh$ of definition \BLUE{of $\sfj$ and its range $\fh$}.
The example $\fh = L^2(\RR)$ with the maps $(\sfj_1 f)(x) := \ol{f(x)}$
and $(\sfj_2 \hat{f})(\xi) := \ol{\hat{f}(\xi)}$ gives a good illustration of
the freedom in the choice of $\sfj$.

Now, suppose that $k \in \ZZ^+$ is a positive integer and $F_1, \ldots,
F_{2k} \in \fh \oplus \fh$ are generalized orbitals. The evenness of
$\rho \in \DM$ is equivalent to the vanishing 
$\Tr_\fF[ \rho \, A^*(F_1) \cdots A^*(F_{2k-1})] =0$ of all
expectation values of monomials of \textit{odd} degree in the
self-dual field operators.  If $\rho \in \DM$ does not preserve the
particle number then expectation values 
$\Tr_\fF[ \rho \, A^*(F_1) \cdots A^*(F_{2k})]$ of monomials of
\textit{even} degree in the self-dual field operators are, in general,
non-vanishing - even if the generalized orbitals are all of the form
$F_j = f_j \oplus 0$, for all $j =1, \ldots, 2k$. 
While the existence of each of these matrix elements is guaranteed by
the boundedness of $A^*(F)$, for any $F \in \fh \oplus \fh$, their
summability requires an extra assumption. To formulate this
we define the subspace
\begin{align} \label{eq-IV.03,01}
\cL_{\NN^k}^1(\fF)
\ := \ 
\big\{ \rho \in \cL^1(\fF) \; \big| \ 
(\NN^{k/2} \rho \NN^{k/2}) \in \cL^1(\fF) \big\}
\ \subseteq \ \cL^1(\fF) \, ,
\end{align}
which is a Banach space with respect to the norm
$\|\rho\|_{\NN^k} :=  
\Tr_\fF\big| (\NN + \bfone)^{k/2} \rho (\NN + \bfone)^{k/2} \big|$. 
We introduce the subset $\DM_{\NN^k} := \DM \cap \cL_{\NN^k}^1(\fF)$ 
of all density matrices $\rho \in \DM$ for which the expectation 
$\la \NN^k \ra_\rho < \infty$ of the $k^{th}$ power of the particle number 
operator is finite.

Given $k \in \ZZ^+$ and a density matrix $\rho \in \DM_{\NN^k}$, we
define its \RED{\textit{reduced generalized $\RED{k}$-particle
  density matrix ($\RED{k}$-gRDM)
  $\RED{\Gamma_\rho^{(k)}} \in \cB\big( (\fh \oplus
  \fh)^{\otimes k} \big)$}} by
\begin{align} \label{eq-IV.04}
\Big\la G_1 \otimes \cdots \otimes G_k \: \Big| \; 
\Gamma_\rho^{(k)} & ( F_1 \otimes \cdots \otimes F_k) \Big\ra
\\ \nonumber
\ := \ &
\Tr_\fF\big[ \rho \: A^*(F_1) \cdots A^*(F_k) \: 
A(G_k) \cdots A(G_1)\, \big] \, ,
\end{align}
where $F_1, \ldots, F_k, G_1, \ldots, G_k \in \fh \oplus \fh$.  We
obtain the \RED{\textit{reduced $\RED{k}$-particle density matrix
  ($\RED{k}$-RDM) $\RED{\gamma_\rho^{(k)}} \in \cB\big(
  \fh^{\otimes k} \big)$}} by restricting the matrix elements on
vectors of the form $g_i \oplus 0$ and $f_j \oplus 0$, that is,
\begin{align} \label{eq-IV.05}
\Big\la g_1 \otimes & \cdots \otimes g_k \: \Big| \; 
\gamma_\rho^{(k)} ( f_1 \otimes \cdots \otimes f_k) \Big\ra
\\ \nonumber
\ := \ &
\bigg\la \begin{pmatrix} g_1 \\ 0 \end{pmatrix} \otimes \cdots \otimes  
\begin{pmatrix} g_k \\ 0 \end{pmatrix} \: \bigg| \; 
\Gamma_\rho^{(k)} \bigg[ 
\begin{pmatrix} f_1 \\ 0 \end{pmatrix} \otimes \cdots \otimes  
\begin{pmatrix} f_k \\ 0 \end{pmatrix} \bigg] \bigg\ra \, .
\end{align}
We observe that in case $\rho$ preserves particle numbers, i.e.,
  $\rho \NN = \NN \rho$, then $\Gamma_\rho^{(k)}$ is entirely determined
by $\gamma_\rho^{(1)}, \gamma_\rho^{(2)}, \ldots, \gamma_\rho^{(k)}$.

The cases $k=1$ and $k=2$ are obviously of special interest.
We first discuss $k=1$ and introduce the
\RED{\textit{pairing operator $\RED{\alpha_\rho}: \fh \to \fh$}} 
corresponding to $\rho$ by 
\begin{align} \label{eq-IV.06}
\big\la g \, \big| \; \alpha_\rho (\sfj f) \, \big\ra 
\ := \
\Tr_\fF\big[ \rho \, c(f) \, c(g) \big] \, ,
\end{align}
noting that $\alpha_\rho$ vanishes if $\rho$ preserves particle
numbers and further that
\begin{align} \label{eq-IV.06,01}
\alpha_\rho^* \ = \ - \sfj \, \alpha_\rho \, \sfj \, .
\end{align}
The pairing operator is convenient for the representation of the
\BLUE{1-gRDM} $\Gamma_\rho^{(1)}: \fh \oplus \fh \to \fh \oplus \fh$ 
given by
\begin{align} \label{eq-IV.07}
\big\la G \, \big| \; \Gamma_\rho^{(1)} \, F \big\ra_{\fh \oplus \fh}
\ = \ &
\Tr_\fF\big[ \rho \, A^*(F) \, A(G) \big] \, ,
\end{align}
for all $F, G \in \fh \oplus \fh$. Viewed as an operator-valued 
$2 \times 2$-matrix acting on vectors $F = f_1 \oplus \sfj f_2$, the
generalized 1-RDM $\Gamma_\rho^{(1)}$ appears as
\begin{align} \label{eq-IV.08}
\Gamma_\rho^{(1)} 
\ = \ &
\begin{pmatrix}
\gamma_\rho^{(1)} & \alpha_\rho \\
\alpha_\rho^* & \bfone - \sfj \, \gamma_\rho^{(1)} \sfj \\
\end{pmatrix} 
\ = \ 
\begin{pmatrix}
\gamma_\rho^{(1)} & \alpha_\rho \\
- \sfj \alpha_\rho \sfj & \bfone - \sfj \, \gamma_\rho^{(1)} \sfj \\
\end{pmatrix} \, ,
\end{align}
where we recall that the \BLUE{1-RDM} $\gamma_\rho^{(1)}$ is given by
$\la g | \, \gamma_\rho^{(1)} f \ra = \Tr_\fF[ \rho c^*(f) c(g)]$.
Eq.~\eqref{eq-IV.08} is equivalent to
\begin{align} \label{eq-IV.09}
\sfJ \, \Gamma_\rho^{(1)} \, \sfJ 
\ = \ &
\bfone \: - \: \Gamma_\rho^{(1)} \, .
\end{align}

Inserting $G = F$ in \eqref{eq-IV.07} and using the anticommutation
relations, it is easily checked that 
$0 \leq \Gamma_\rho^{(1)} \leq \bfone$ holds true which, in turn, 
is equivalent to
\begin{align} \label{eq-IV.10}
\begin{pmatrix}
\gamma_\rho^{(1)} - (\gamma_\rho^{(1)})^2 - \alpha_\rho \alpha_\rho^* 
& \gamma_\rho^{(1)} \alpha_\rho - \alpha_\rho \sfj \, \gamma_\rho^{(1)} \sfj
\\[1ex]
[\gamma_\rho^{(1)} \alpha_\rho - 
\alpha_\rho \sfj \, \gamma_\rho^{(1)} \sfj ]^* 
& \sfj \, [ \gamma_\rho^{(1)} - (\gamma_\rho^{(1)})^2 
- \alpha_\rho \alpha_\rho^* ] \sfj \\
\end{pmatrix} 
\ = \
\Gamma_\rho^{(1)} \: - \: \big( \Gamma_\rho^{(1)} \big)^2 
\ \geq \ 0 \, .
\end{align}
Note that this yields $\gamma_\rho^{(1)} - (\gamma_\rho^{(1)})^2 \geq 0$
and hence $0 \leq \gamma_\rho^{(1)} \leq \bfone$, as asserted in
Lemma~\ref{lem-II.01}~(ii). 

Furthermore, if $\rho \in \DM_{\NN}$ has finite 
particle number expectation then
\eqref{eq-IV.10} implies that the pairing operator 
$\alpha_\rho \in \cL^2(\fh)$ is Hilbert-Schmidt, with
$\Tr_\fh[\alpha_\rho^* \alpha_\rho] \leq \Tr_\fh[\gamma_\rho^{(1)} -
(\gamma_\rho^{(1)})^2]$, and that 
$\Gamma_\rho^{(1)} - (\Gamma_\rho^{(1)})^2 \in \cL^1(\fh \oplus \fh)$
is trace-class. In particular, $\Gamma_\rho^{(1)} - (\Gamma_\rho^{(1)})^2$
and hence also $\Gamma_\rho^{(1)}$ admits an expansion of the form
$\Gamma_\rho^{(1)} = 
\sum_{i=1}^\infty \tlambda_i \, | F_i \ra\la F_i |$,
where $\tlambda_i \in [0,1]$ are its eigenvalues and 
$\{F_i\}_{i=1}^\infty \subseteq \fh \oplus \fh$ is an orthonormal basis
of eigenvectors $F_i = f_i' \oplus \sfj f_i''$ of $\Gamma_\rho^{(1)}$. 
The invariance $\sfJ \Gamma_\rho^{(1)} \sfJ = \bfone - \Gamma_\rho^{(1)}$
implies that the eigenvalues and corresponding eigenvectors
come in pairs $\lambda_\ell$, $F_\ell$ and $1-\lambda_\ell$, $\sfJ F_\ell$. 
After changing the order of the eigenvalues, if necessary,
we obtain 
\begin{align} \label{eq-IV.11}
\Gamma_\rho^{(1)} \ = \ 
\sum_{\ell=1}^\infty \BLUE{\Big\{} \lambda_\ell \: | F_\ell \ra\la F_\ell | 
\: + \: (1-\lambda_\ell) \: | \sfJ F_\ell \ra\la \sfJ F_\ell | 
\BLUE{\Big\}} \, ,
\end{align}
where $F_\ell = f_\ell' \oplus \sfj f_\ell''$ and 
$\{F_\ell, \sfJ F_\ell\}_{\ell=1}^\infty \subseteq \fh \oplus \fh$ 
is an orthonormal basis. If, additionally, $\rho$ is
particle-number preserving and so $\alpha_\rho \equiv 0$, then  
\begin{align} \label{eq-IV.12}
\Gamma_\rho^{(1)} 
\ = \ &
\gamma_\rho^{(1)} \oplus \big( \bfone - \sfj \, \gamma_\rho^{(1)} \sfj \big) 
\\[1ex] \nonumber 
\ = \ &
\sum_{\ell=1}^\infty \BLUE{\Big\{} 
\lambda_\ell \: | f_\ell \oplus 0 \ra\la f_\ell  \oplus 0 | \: + \:
(1-\lambda_\ell) \: | 0 \oplus \sfj f_\ell \ra\la 0 \oplus \sfj f_\ell | 
\BLUE{\Big\}} \, ,
\end{align}
where $\lambda_\ell$ are the eigenvalues of $\gamma_\rho^{(1)}$ and
$\{f_\ell\}_{\ell=1}^\infty \subseteq \fh$ is an orthonormal basis of
its eigenvectors. Since 
$\sum_{\ell=1}^\infty \lambda_\ell = \Tr_\fh[\gamma_\rho^{(1)}]<\infty$, 
the sequence of eigenvalues including their multiplicities is summable. 

\paragraph*{\GREEN{Bogoliubov} Transformations}
For a density matrix $\rho \in \DM_{\NN}$ of finite 
particle number expectation the block-diagonal form \eqref{eq-IV.12} 
of \BLUE{its 1-gRDM} $\Gamma_\rho^{(1)}$ can always be obtained by conjugation 
$\UU_W \rho \UU_W^*$ of $\rho$ by a (unitary) \GREEN{Bogoliubov} transformation
$\UU_W \in U(\fF)$ on Fock space corresponding to a 
\GREEN{\textit{Bogoliubov linear map}}, i.e., a unitary $W \in U(\fh
\oplus \fh)$ on $\fh \oplus \fh$, which additionally obeys $\sfJ W = W
\sfJ$. The latter condition and the unitarity precisely ensure
\BLUE{$\UU_W \Om \in \fF$ and} that the CAR \eqref{eq-IV.03} are
preserved under these transformations,
\begin{align} \label{eq-IV.13}
A(W F) \ = \ & A^*( \sfJ W F ) \ = \ A^*( W \sfJ F )  
\quad \text{and} 
\\[1ex] \label{eq-IV.14}
\big\{ A(W F) , A^*(W & G) \big\}
\ = \
\big\la W F \big| W G \big\ra_{\fh \oplus \fh} 
\ = \ 
\big\la F \big| G \big\ra_{\fh \oplus \fh} \, ,
\end{align}
The \GREEN{Bogoliubov} linear maps obviously form a subgroup
\begin{align} \label{eq-IV.15}
\Bog_{\fh \oplus \fh} 
\ := \ &
\big\{ W \in U(\fh \oplus \fh) \; \big| 
\ \ \sfJ \, W \ = \ W \, \sfJ \; \big\} 
\end{align}
of $U(\fh \oplus \fh)$. Expressing $W$ as a $2 \times 2$-matrix of
operators, the \GREEN{Bogoliubov} linear maps can be alternatively
characterized as
\begin{align} \label{eq-IV.16}
\Bog_{\fh \oplus \fh} 
\ = \ &
\bigg\{ \begin{pmatrix} 
u & \sfj v \sfj \\ 
v & \sfj u \sfj \\
\end{pmatrix} \; \in \; U(\fh \oplus \fh) \bigg| 
\ \Tr_\fh\big[ v^* v \big] \; < \; \infty \bigg\} \, ,
\end{align}
where the condition that $v$ is of Hilbert-Schmidt class, 
$\Tr_\fh[ v^* v ] < \infty$, is known as the \RED{\textit{Shale-Stinespring
  condition}}. Each \GREEN{Bogoliubov} linear map $W \in \Bog_{\fh \oplus \fh}$ 
is unitarily implementable on Fock space which means that there
exists a unitary $\UU_W \in U(\fF)$ such that, for all 
$F \in \fh \oplus \fh$, 
\begin{align} \label{eq-IV.17}
\UU_W \, A^*(F) \, \UU_W^* \ = \ & A^*( W F) 
\end{align}
and, in fact, $W \mapsto \UU_W$ is a bijection 
$\Bog_{\fh \oplus \fh} \to \Bog_\fF$, where
\begin{align} \label{eq-IV.18}
& \Bog_\fF \ := \ 
\\ \nonumber
& \quad \Big\{ \UU \in U(\fF) \; \Big| 
\ \exists \, V \in \cB(\fh \oplus \fh)
\ \forall \, F \in \fh \oplus \fh: \
\UU \, A^*(F) \, \UU^* \; = \; A^*(V F) \Big\} \, 
\end{align}
is the subgroup $\RED{\Bog_\fF} \subseteq U(\fF)$ of
\GREEN{\textit{Bogoliubov transformations}}. The Shale-Stinespring condition
ensures, that the vacuum vector remains in $\fF$ under the application
of $\UU_W$, and the transformed creation and annihilation operators
$d^*(f) := \UU_W c^*(f) \UU_W^*$ and $d(f)$ constitute another Fock
representation of the CAR with $\UU_W \Om \in \fF$ as the new vacuum
vector.

We return to the 1-gRDM $\Gamma_\rho^{(1)}$ of
a density matrix $\rho \in \DM_{\NN}$ of finite particle 
number expectation. These assume the form \eqref{eq-IV.08} with
nonvanishing pairing operator $\alpha_\rho$ unless $\rho$ preserves
particle numbers. In an orthonormal basis 
$\{ F_\ell , \sfJ F_\ell \} \subseteq \fh \oplus \fh$ of eigenvectors
with eigenvalues $\lambda_\ell$ and $1-\lambda_\ell$, respectively,
$\Gamma_\rho^{(1)}$ can be represented as in
\eqref{eq-IV.11}. Starting from this one can construct a \GREEN{Bogoliubov}
linear map $W \in \Bog_{\fh \oplus \fh}$ such that
\begin{align} \label{eq-IV.19}
W^* \, \Gamma_\rho^{(1)} \, W 
\ = \ &
\sum_{\ell=1}^\infty \Big\{ 
\lambda_\ell \: | f_\ell \oplus 0 \ra\la f_\ell  \oplus 0 | \: + \:
(1-\lambda_\ell) \: | 0 \oplus \sfj f_\ell \ra\la 0 \oplus \sfj f_\ell | 
\Big\} \, .
\end{align}
Since, for all $F, G \in \fh \oplus \fh$,
\begin{align} \label{eq-IV.20}
\Tr_\fF\big[ \rho \, A^*(W F) \, A(W G) \big] 
\ = \ 
\Tr_\fF\big[ \UU_W^* \, \rho \, \UU_W \, A^*(F) \, A(G) \big] \, ,
\end{align}
we obtain that 
\begin{align} \label{eq-IV.21}
\Gamma_{\UU_W^* \rho \UU_W}^{(1)} 
\ = \ &
W^* \, \Gamma_\rho^{(1)} \, W 
\\[1ex] \nonumber
\ = \ &
\sum_{\ell=1}^\infty \Big\{ 
\lambda_\ell \: | f_\ell \oplus 0 \ra\la f_\ell  \oplus 0 | \: + \:
(1-\lambda_\ell) \: | 0 \oplus \sfj f_\ell \ra\la 0 \oplus \sfj f_\ell | 
\Big\} \, .
\end{align}
In other words, the pairing operator $\alpha_{\UU_W^* \rho \UU_W} =0$
of the transformed density matrix $\UU_W^* \rho \UU_W$ vanishes
and $\gamma_{\UU_W^* \rho \UU_W}^{(1)} = \sum_{\ell =1}^\infty \lambda_\ell
|f_\ell \ra\la f_\ell|$ where $\{ f_\ell \} \in \fh$ is an orthonormal
basis and $\lambda_\ell \in [0,1]$. Note that the vanishing
$\alpha_{\UU_W^* \rho \UU_W} =0$ of the pairing operator alone does not
imply that $\UU_W^* \rho \UU_W$ is particle-number preserving.
Further note that if
$W = \big( \begin{smallmatrix} 
u & \sfj v \sfj \\ 
v & \sfj u \sfj 
\end{smallmatrix} \big) \in \Bog_{\fh \oplus \fh}$
then
\begin{align} \label{eq-IV.22}
0 \ \leq \ 
\gamma_{\UU_W^* \rho \UU_W}^{(1)} 
\ = \ &
u^* \gamma_\rho^{(1)} u + v^* \big( \bfone - \sfj \gamma_\rho^{(1)} \sfj \big) v
+ v^* \alpha_\rho u + u^* \alpha_\rho v
\nonumber \\[1ex]
\ \leq \ &
u^* \gamma_\rho^{(1)} u + v^* v
+ v^* \alpha_\rho u + u^* \alpha_\rho v \, ,
\end{align}
from which we conclude that the transformed density matrix 
$\UU_W^* \rho \UU_W$ has finite particle number expectation, as well,
since
\begin{align} \label{eq-IV.23} 
\| \gamma_{\UU_W^* \rho \UU_W}^{(1)} \|_{\cL^1}
\ \leq \ &
\| \gamma_\rho^{(1)} \|_{\cL^1}
+ \| v \|_{\cL^2}^2 + 2 \| v \|_{\cL^2} \, \| \alpha_\rho \|_{\cL^2}
\ < \ \infty \, .
\end{align}
Inspired by these properties, we define by 
\begin{align} \label{eq-IV.24}
& \genonepdm 
\ := \ 
\\ \nonumber
& \bigg\{ \Gamma^{(1)} = \big(\begin{smallmatrix}
\gamma^{(1)} & \alpha \\
\alpha^* & \bfone - \sfj \, \gamma^{(1)} \sfj \\
\end{smallmatrix}\big) 
\; \in \; \cB(\fh \oplus \fh) \ \bigg| \ 
\Gamma^{(1)} = \sfJ ( \bfone - \Gamma^{(1)}) \sfJ \geq 0 \, , \
\gamma^{(1)} \in \cL^1(\fh)
\bigg\} 
\end{align}
the set of \RED{\textit{generalized one-particle density matrices 1-gpdm}}
and by
\begin{align} \label{eq-IV.25}
\onepdm  
\ := \ & 
\Big\{ \gamma^{(1)} \in \cL^1(\fh) \: \Big| \ 
0 \leq \gamma^{(1)} \leq \bfone \Big\} 
\end{align}
the set of \RED{\textit{one-particle density matrices (1-pdm)}}.

\paragraph*{Representability of 1-gpdm} 
We have just seen that any 1-gRDM of finite particle-number
expectation value \BLUE{necessarily} is a 1-gpdm in
$\genonepdm$. Representability asks for \BLUE{sufficient} conditions
for this relation. That is, a 1-gpdm $\Gamma^{(1)} \in \genonepdm$ is called
\RED{\textit{representable}}, if there exists a density matrix $\rho \in
\DM$ whose reduced generalized one-particle density matrix
$\Gamma_\rho^{(1)}$ coincides with $\Gamma^{(1)}$, i.e., if
$\Gamma^{(1)} = \Gamma_\rho^{(1)}$.

The following theorem gives an affirmative answer
to question of representability of generalized 1-pdm.
%
\begin{theorem} \label{thm-IV.01} Every generalized one-particle 
density matrix $\Gamma^{(1)} \in \genonepdm$ is representable
by a density matrix of finite particle number expectation value.
\begin{proof} Given $\Gamma^{(1)} \in \genonepdm$ we can find a
\GREEN{Bogoliubov} linear map $W \in \Bog_{\fh \oplus \fh}$ such that 
\begin{align} \label{eq-IV.26}
W^* \, \Gamma^{(1)} \, W 
\ = \ &
\begin{pmatrix}
\gamma & 0 \\
0 & \bfone - \sfj \, \gamma \sfj \\
\end{pmatrix} 
\quad \text{and} \quad
\gamma \ = \ \sum_{\ell=1}^\infty \lambda_\ell \, | f_\ell \ra\la f_\ell | 
\end{align}
assumes the form \eqref{eq-IV.19}. Here, 
$\{ f_\ell \}_{\ell=1}^\infty \subseteq \fh$ is an orthonormal basis
of eigenvectors of $\gamma$ with corresponding eigenvalues
$\lambda_\ell \in [0,1]$, which we assume w.l.o.g.\ to be
arranged in descending order,
$1 \geq \lambda_1 \geq \lambda_2 \geq \ldots \geq 0$.
More specifically, we have that
$1 = \lambda_1 = \ldots = \lambda_{K-1} > 
\lambda_K \geq \ldots \geq \lambda_L > \lambda_{L+1} = 
\lambda_{L+2} = \ldots = 0$, for unique 
$K \leq \Tr_\fh[\gamma] < \infty$
and $L \in \ZZ^+ \cup \{\infty\}$. Note that, for $K \leq \ell \leq L$, 
the eigenvalues $\lambda_\ell \in [\lambda_L, \lambda_K] \subseteq (0,1)$
are away from $0$ and $1$, and 
$\mu_\ell := \ln(1-\lambda_\ell) - \ln(\lambda_\ell) \in \RR$ exists.
Setting $n_k := c^*(f_k) \, c(f_k)$, for all $k \in \ZZ^+$, 
and $\PP_1 := n_1 \, n_2 \cdots n_{K-1}$, we define
\begin{align} \label{eq-IV.27}
\hh_0 \ := \ \sum_{\ell=K}^L \mu_\ell \: n_\ell 
\, , \ \ 
Z_0 \ := \ \Tr_\fF[e^{-\hh_0}] \, , 
\ \ \text{and} \ \ 
\rho_0 \ := \ \PP_1 \, Z_0^{-1} \: \exp[-\hh_0] \, .
\end{align}
Note that an orthonormal basis of $\fF$ of eigenvectors
of $n_k$ with eigenvalues $\nu_k \in \{0,1\}$ 
is given by 
$\Psi_{\unu} := \prod_{\ell=1}^\infty [c^*(f_\ell)]^{\nu_\ell} \Om$,
where $\unu = (\nu_\ell)_{\ell=1}^\infty \in \{0,1\}^{\ZZ^+}$ runs
through all sequences of occupation numbers $\nu_\ell \in \{0,1\}$
of finite sum $|A(\unu)| < \infty$, with
$A(\unu) := \{ \ell \in \ZZ^+ | \nu_\ell = 1 \} \subseteq \ZZ^+$.
That is, $n_k \Psi_{\unu} = \nu_k \Psi_{\unu}$, for any $k \in \ZZ^+$.
Hence
\begin{align} \label{eq-IV.28}
Z_0 \ = \ 
\sum_{\unu: |A(\unu)| < \infty} 
\big\la \Psi_{\unu} \, \big| \; e^{-\hh_0} \Psi_{\unu} \big\ra
\ = \ 
\prod_{\ell=K}^L \big( 1 + e^{-\mu_\ell} \big) \ < \ \infty \, ,
\end{align}
since $\sum_{\ell=1}^\infty e^{-\mu_\ell} = 
\sum_{\ell=K}^L (1-\lambda_\ell)^{-1} \lambda_\ell \leq
(1-\lambda_K)^{-1} \sum_{\ell=1}^\infty \lambda_\ell < \infty$.  
It follows that $\rho_0 \in \DM$ is a density matrix, which
is obviously particle-number preserving and, therefore, has
vanishing pairing operator $\alpha_{\rho_0} = 0$. Moreover,
if $\max\{k,\ell\} \geq K$ then
\begin{align} \label{eq-IV.29}
\la f_\ell \, | \; \gamma_{\rho_0}^{(1)} f_k \ra 
\ = \
\frac{\Tr_\fF[ e^{-\hh_0} \, c^*(f_k) \, c(f_\ell) ]}{Z_0}
\ = \
\frac{\delta_{k,\ell} \; e^{-\mu_k}}{1 + e^{-\mu_k}}
\ = \
\delta_{k,\ell} \; \lambda_k \, ,
\end{align}
while, for $\min\{k,\ell\} \leq K$, we observe that
$\la f_\ell | \gamma_{\rho_0}^{(1)} f_k \ra = \delta_{k,\ell} = 
\delta_{k,\ell} \lambda_k$, as well. This implies that 
$W^* \Gamma^{(1)} W = \Gamma_{\rho_0}^{(1)}$ and thus 
\begin{align} \label{eq-IV.30}
\Gamma^{(1)} 
\ = \ &
W \, \Gamma_{\rho_0}^{(1)} \, W^*  
\ = \ 
\Gamma_{\UU_W \, \rho_0 \, \UU_W^*}^{(1)} \, .
\end{align}
Since $\rho_0 \in \DM$ is a density matrix, so is 
$\UU_W \, \rho_0 \, \UU_W^* \in \DM$.
\end{proof}
\end{theorem}

\paragraph*{$\RED{N}$-Representability of 1-pdm} 
Similar to the notion of representability of a generalized 1-pdm,
we call a 1-pdm $\gamma^{(1)} \in \onepdm$ with 
$\Tr[\gamma^{(1)}] =N \in \ZZ^+$
\RED{\textit{$\RED{N}$-representable}}, if 
there exists an $N$-particle density matrix $\rho \in \DM^{(N)}$
such that $\gamma^{(1)} = \gamma_\rho^{(1)}$.

The $N$-representability of any 1-pdm has actually been proved
by Lieb in \cite{Lieb1981a}, although this had not been its main purpose.
%
\begin{theorem} \label{thm-IV.02} Let $N \in \ZZ^+$ with $N \geq 2$
and $\gamma^{(1)} \in \onepdm$ \RED{be} a one-particle density matrix 
of particle number expectation $\Tr[\gamma^{(1)}] = N$.
Then $\gamma^{(1)}$ is $N$-representable.
\begin{proof} Given $\gamma^{(1)}$, the $N$-particle density matrix 
$\rho_\av \in \DM^{(N)}$ in \eqref{eq-IV.01} fulfills
$\gamma^{(1)} = \gamma_{\rho_\av}^{(1)}$.
\end{proof}
\end{theorem}

\paragraph*{Representability of generalized 2-pdm} 
Let $N \in \ZZ^+$ with $N \geq 2$. As proven in Theorems~\ref{thm-IV.01} 
and \ref{thm-IV.02} above, the maps 
$\DM_{\NN} \to \genonepdm$, $\rho \mapsto \Gamma_\rho^{(1)}$ 
and $\DM^{(N)} \to \{ \gamma \in \onepdm | \Tr[\gamma] = N\}$, 
$\rho \mapsto \gamma_\rho^{(1)}$ are bijections. The simple
characterizations of the sets $\genonepdm$ and $\onepdm$ 
\RED{are} an encouraging sign that the extension of the notion of
representability to reduced generalized $k$-pdm for $k \geq 2$
leads to similarly simple characterizations. 

Following this sign, we call a pair 
$(\Gamma^{(1)}, \Gamma^{(2)}) \in \cB(\fh^2) \times \cB(\fh^2 \otimes \fh^2)$ 
of bounded positive operators \RED{\textit{representable}}, if 
$\Gamma^{(1)} = \Gamma_\rho^{(1)}$ and $\Gamma^{(2)} = \Gamma_\rho^{(2)}$,
for some density matrix $\rho \in \DM_{\la \NN^2 \ra < \infty}$ of
finite particle number variance\, where $\fh^2 := \fh \oplus \fh$. 

Somewhat more restrictive, we call a pair 
$(\gamma^{(1)}, \gamma^{(2)}) \in \cB(\fh) \times \cB(\fh \otimes \fh)$ 
of bounded positive operators \RED{\textit{representable}}, if 
$\gamma^{(1)} = \gamma_\rho^{(1)}$ and $\gamma^{(2)} = \gamma_\rho^{(2)}$,
for some particle-number preserving density matrix 
$\rho \in \DM_{\la \NN^2 \ra < \infty}$ of
finite particle number variance. 
If $\rho$ can additionally be chosen to be an $N$-particle density 
matrix then $(\gamma^{(1)}, \gamma^{(2)})$, respectively, is called
\RED{\textit{$\RED{N}$-representable}}. Note that necessarily 
\BBBLUE{$\gamma^{(1)}$ results from $\gamma^{(2)}$ by taking a partial trace
[see \eqref{eq-II.25}] and} $N = \Tr[\gamma^{(1)}]$ in this case.

With these definitions we obtain new characterizations of the
total and the $N$-particle ground state energies as 
\begin{align} \label{eq-IV.31}
E_\gs \ = \ &
\inf\Big\{ 
\cE_Q\big( \gamma^{(1)}, \gamma^{(2)} \big) \: \Big|
\ ( h \, \gamma^{(1)} ) \in \cL^1(\fh) \, , \ 
\\ \nonumber & \qquad\qquad
\text{$( \gamma^{(1)}, \gamma^{(2)} ) 
\in \cB(\fh) \times \cB(\fh \otimes \fh)$ is representable} \Big\} \, ,
\\[1ex] \label{eq-IV.32}
E_\gs(N) \ = \ &
\inf\Big\{ 
\cE_Q\big( \gamma^{(1)}, \gamma^{(2)} \big) \: \Big|
\ ( h \, \gamma^{(1)} ) \in \cL^1(\fh) \, , \ 
\\ \nonumber & \qquad\qquad
\text{$( \gamma^{(1)}, \gamma^{(2)} ) 
\in \cB(\fh) \times \cB(\fh \otimes \fh)$ is $N$-representable} \Big\} \, .
\end{align}
This characterization of the ground state energy was first given by
Coleman \cite{Coleman1963}, following a remark \RED{by} Coulson
\cite{Coulson1960}. It seems to yield a drastic simplification of the
task of determining ground state energies and ground states of
many-fermion systems, as the number of variables of the problem is
reduced from $N$ to $4$.  This is, however, too optimistic because the
problem of restricting the variation in \eqref{eq-IV.31} and
\eqref{eq-IV.32} to representable, respectively $N$-representable, pairs
$(\gamma^{(1)}, \gamma^{(2)})$ is, perhaps, as difficult as solving
the corresponding Schrödinger equation on Fock space altogether.

The requirement that the density matrix from which 
$(\gamma^{(1)}, \gamma^{(2)})$ derives is particle number preserving
or even an $N$-particle density matrix adds considerably to the degree
of difficulty of the problem, as is seen when comparing the proofs of
Theorems~\ref{thm-IV.01} and \ref{thm-IV.02} in case that $k=1$.  A
characterization of the representability of 
$(\Gamma^{(1)}, \Gamma^{(2)}) \in \cB(\fh^2) \times \cB(\fh^2 \otimes
\fh^2)$ \RED{would be} great progress.

Nevertheless, we now focus on particle number preserving density
matrices $\rho$ for which the reduced generalized 1-pdm
$(\Gamma_\rho^{(1)}, \Gamma_\rho^{(2)})$ are completely determined by
the 1-RDM $(\gamma_\rho^{(1)}, \gamma_\rho^{(2)})$. The difficulty
described above has lead to what is known as the
\RED{\textit{representability problem}} of quantum chemistry: \textit{
  Specify a condition $A: \cL^1(\fh) \times \cL^1(\fh \otimes \fh) \to
  \{\text{true}, \text{false}\}$ such that $( \gamma^{(1)},
  \gamma^{(2)} )$ is representable if $A( \gamma^{(1)}, \gamma^{(2)} )
  = \text{true}$.}  
The representability problem is \RED{still considered open today} (at
least by those who do not accept tautologies as its solution). It is
known to be a hard problem in the sense of QMA complexity in computer
science, \BBBLUE{as demonstrated by Liu, Christandl, and Verstraete in
  \cite{LiuChristandlVerstraete2007}. An overview on questions of
  reduced density matrices and their representability is given by
  Coleman and Yukalov in \cite{ColemanYukalov2000}.}

\paragraph*{GPQ Condition and $\RED{T_{1;2}}$ Condition} While the
representability problem\BLUE{, which} is about the specification of a
\textit{sufficient} condition for the representability of a pair $(
\gamma^{(1)}, \gamma^{(2)} )$, \BLUE{remains open}, research on
conditions reduced one- and two-particle density matrices
\textit{necessarily} fulfill has been more successful in the past.
Namely, if a condition $B: \cL^1(\fh) \times \cL^1(\fh \otimes \fh)
\to \{\text{true}, \text{false}\}$ is such that $B( \gamma_\rho^{(1)},
\gamma_\rho^{(2)} ) = \text{true}$, for any density matrix $\rho \in
\DM$ then it is immediate that
\begin{align} \label{eq-IV.33}
E_\gs \ \geq \ &
\inf\Big\{ 
\cE_Q\big( \gamma^{(1)}, \gamma^{(2)} \big) \: \Big|
\\ \nonumber & \qquad\qquad
\ ( h \, \gamma^{(1)} ) \in \cL^1(\fh) \, , \ 
B( \gamma^{(1)}, \gamma^{(2)} ) = \text{true} \Big\} \, ,
\\[1ex] \label{eq-IV.34}
E_\gs(N) \ \geq \ &
\inf\Big\{ 
\cE_Q\big( \gamma^{(1)}, \gamma^{(2)} \big) \: \Big|
\\ \nonumber & \qquad\qquad
\ ( h \, \gamma^{(1)} ) \in \cL^1(\fh) \, , \ \Tr[\gamma^{(1)}] = N \, , \ 
B( \gamma^{(1)}, \gamma^{(2)} ) = \text{true} \, \Big\} \, .
\end{align}
In practise, $B$ is not a single condition but a list of conditions
that $\gamma^{(1)}$ and $\gamma^{(2)}$ ought to fulfill, and
\GREEN{Conditions~(i) and (ii)} in Lemma~\ref{lem-II.01} are always
part of this list. That is, it is understood that $\gamma^{(1)} \in
\onepdm$ is a 1-pdm and obeys $0 \leq \gamma^{(1)} \leq \bfone_\fh$
and $\Tr[\gamma^{(1)}] < \infty$.  Theorems~\ref{thm-IV.01} and
\ref{thm-IV.02} ensure that there are not more conditions on
$\gamma^{(1)}$ alone, that do not involve $\gamma^{(2)}$.

Almost sixty years ago Coleman \cite{Coleman1963} and Garrod and
Percus \cite{GarrodPercus1964} specified three conditions, which a
representable pair $(\gamma^{(1)}, \gamma^{(2)})$ of a one- and
two-particle density matrix necessarily fulfill. These three
conditions were orginally called ``G'', ``P'', and ``Q'',
respectively, but we refer to them as a single condition 
which we call the \RED{\textit{GPQ condition}}. We apply the scheme
described in \eqref{eq-IV.33} and \eqref{eq-IV.34} above and
introduce
\begin{align} \label{eq-IV.35}
E_\GPQ(N) \ := \ &
\inf\big\{ 
\cE_Q( \gamma^{(1)}, \gamma^{(2)} ) \: \big| 
\ ( h \, \gamma^{(1)} ) \in \cL^1(\fh) \, , \ 
\\ \nonumber & \qquad\qquad
\Tr[\gamma^{(1)}] = N \, , \ 
\text{$( \gamma^{(1)}, \gamma^{(2)} )$ fulfills GPQ} \big\} \, ,
\end{align}
observing that $E_\gs(N) \geq E_\GPQ(N)$.  In
\cite{BachKnoerrMenge2012}, Knörr, Menge, and the author considered
self-adjoint, but not necessarily positive, trace-class operators
$\rho = \rho^* \in \cL^1(\fF)$ obeying $\Tr_\fF\big(|\rho|^{1/2} \NN^2
|\rho|^{1/2} \big) < \infty$.  It is easy to see that, for these
$\rho$, the operators $\Gamma_\rho^{(2)}$, given by \eqref{eq-IV.04},
define trace-class operators on $\fh^2 \otimes \fh^2$. In
\cite{BachKnoerrMenge2012}, the GPQ condition was proven to be is
equivalent to the positivity of $\Gamma_\rho^{(2)} \geq 0$ on $\fh^2
\otimes \fh^2$. Furthermore, it was shown in
\cite{BachKnoerrMenge2012} that the GPQ condition implies the fermion
correlation inequality
\begin{align} \label{eq-IV.36}
\Tr_{\fh \otimes \fh}\big[ (P \otimes P) \gamma^{(2)} \big] 
\ \geq \ & 
\Tr_{\fh \otimes \fh}\big[ (P \otimes P) (\bfone - \Ex) 
(\gamma^{(1)} \otimes \gamma^{(1)}) \big] 
\\[1ex] \nonumber
& - 
\Tr_\fh[ P \gamma^{(1)} ] \; \min\Big\{ 1 \; , \; 9 \, 
\Tr_\fh\big[ P \big( \gamma^{(1)} - (\gamma^{(1)})^2 \big)^{1/2} \big] 
\Big\} \, ,
\end{align}
where $P = P^* = P^2 \in \cB(\fh)$ is an arbitrary orthogonal
projection. This inequality is the key input for the proof in
\cite{Bach1992, BachKnoerrMenge2012} that, for large Coulomb systems,
the difference of the Hartree--Fock energy and $E_\GPQ(N)$ is bounded
by $o(Z^{5/3})$, which implies that the accuracy of the Hartree--Fock
approximation is at least as good,
\begin{align} \label{eq-IV.37}
0 \ \leq \ 
E_\HF(Z) - E_\gs(N) 
\ \leq \ 
E_\HF(Z) - E_\GPQ(N) 
\ \leq \ o(Z^{(5/3)}) \, .
\end{align}
\BLUE{A similar inequality was established
and then applied to Fermi Jellium (described below) 
by Graf and Solovej in \cite{GrafSolovej1994}.}
Since the exchange term is in magnitude greater than a universal
multiple of $Z^{5/3}$, see \eqref{eq-I.09,02}-\eqref{eq-I.09,03},
Eq.~\eqref{eq-IV.37} proves that the accuracy of the Hartree--Fock
approximation is better than the smallest contribution to the
Hartree--Fock energy. 

In \cite{Erdahl1978a, Erdahl1978b}, Erdahl found additional
representability conditions he called $T_1$ and $T_2$. We refer to
these as a single condition, the \RED{\textit{$\RED{T_{1;2}}$
  condition}}. It arises from observables of the form $Q_4 := P_3^*
P_3 + P_3 P_3^*$, where $P_3$ is any polynomial in the self-dual field
operators of degree three. Obviously, $Q_4$ is a nonnegative
operator. Moreover, while both $P_3^* P_3$ and $P_3 P_3^*$ are
polynomials of degree six, their sum $Q_4$ is an anticommutator and
hence a polynomial of degree four or less. Thus, $\Tr_\fF[ \rho \, Q_4
] \geq 0$ yields a condition the pair of reduced generalized 1-pdm and
2-pdm $(\Gamma_\rho^{(1)}, \Gamma_\rho^{(2)})$ corresponding to $\rho$
necessarily fulfills. We introduce
\begin{align} \label{eq-IV.38}
E_\GPQT(N) \ := \ &
\inf\big\{ 
\cE_Q( \gamma^{(1)}, \gamma^{(2)} ) \: \big|
\ ( h \, \gamma^{(1)} ) \in \cL^1(\fh) \, , \ 
\\ \nonumber & \qquad\qquad
\Tr[\gamma^{(1)}] = N \, , \ 
\text{$( \gamma^{(1)}, \gamma^{(2)} )$ fulfills GPQ and $T_{1;2}$} 
\big\} \, .
\end{align}
Erdahl's theoretical discovery came into focus of quantum chemists
some two decades later, when numerical simulations demonstrated, that,
in test cases with small $N$, the accuracy of $E_\GPQT(N)$ is
comparable to the accuray of \textit{full CI (configuration
  interaction)} computations, i.e., the full solution of the $N$
electron Schr{\"o}dinger equation (projected onto a finite dimensional
subspace, as part of the Galerkin approximation). These were carried
out, e.g., by \RED{Mazziotti and Erdahl} in \cite{MazziottiErdahl2001},
Zhao, Braams, Fukuda, Overton, and Percus in
\cite{ZhaoBraamsFukudaOvertonPercus2004}, Cances, Lewin, and Stoltz in
\cite{CancesLewinStoltz2006}, Braams, Percus, and Zhao in
\cite{BraamsPercusZhao2007}, and Naftchi-Ardebili, Hau, and Mazziotti
in \cite{Naftchi-ArdebiliHauMazziotti2011}.

\newpage
\section{Quadratic Hamiltonians and \\
Quasifree Density Matrices} 
\label{sec-V}
%
\paragraph*{Quadratic Hamiltonians:} We return to the definition of
\GREEN{Bogoliubov} linear maps $\Bog_{\fh \oplus \fh}$ and \GREEN{Bogoliubov}
transformations $\Bog_\fF$. The former consists of unitary linear maps
$W \in U(\fh \oplus \fh)$ on $\fh \oplus \fh$, which additionally
obeys $\sfJ W = W \sfJ$, the latter are unitary operators 
$\UU_W \in U(\fF)$ obeying 
\begin{align} \label{eq-V.01} 
\forall \, F \in \fh \oplus \fh: \qquad
\UU_W \, A^*(F) \, \UU_W^* \ = \ A^*( W F) \, , 
\end{align}
and the map $\Bog_{\fh \oplus \fh} \ni W \mapsto \UU_W \in \Bog_\fF$
is a group isomorphism.

Next, we define the second quantization $\QQ(T) \in \cB(\cD(N); \fF_f)$ 
of a bounded operator $T = T^* = 
\big( \begin{smallmatrix} a & b \\ b^* & 0 \end{smallmatrix} \big) 
\in \cB[\fh \oplus \fh]$, with $a = a^*$ and $b = - \sfj b^* \sfj$ by
\begin{align} \label{eq-V.02} 
\QQ(T) \ := \ 
\sum_{i,j=1}^\infty \la F_i | \: T F_j \ra \: A^*(F_i) \, A(F_j) \, , 
\end{align}
where $\{F_i\}_{i = 1}^\infty \subseteq \fh \oplus \fh$ is an
orthonormal basis. The definition of $\QQ(T)$ is independent of the
choice of this orthonormal basis. Under the assumption that $b \in
\cL^2(\fh)$ is a Hilbert-Schmidt operator and $a \geq 0$ is
nonnegative, $\QQ(T)$ is self-adjoint and semibounded on the domain of
the particle number operator. (Generally, a relative bound in form of
the Hilbert-Schmidt property of $a^{-1/2} b a^{-1/2}$ should be
sufficient, as this was shown to hold true for boson systems by
  Nam, Napiorkowski, and Solovej in
\cite{NamNapiorkowskiSolovej2016}.) We refer to $\RED{\QQ(T)}$
as the \RED{\textit{quadratic Hamiltonian corresponding to
  $\RED{T}$}} because it is of degree two in the self-dual field
operators. An explicit computation (on finite vectors and then
extension by continuity) yields
\begin{align} \label{eq-V.03} 
[ \QQ(T) \, , \, A^*(F) ] \ = \  A^*\big( \hT F \big) \, , 
\end{align}
which implies that
\begin{align} \label{eq-V.04} 
e^{i \QQ(T)} \, A^*(F) \, e^{-i \QQ(T)} 
\ = \  
A^*\big( e^{i\hT} F \big) \, ,
\end{align}
for any $F \in \fh \oplus \fh$, where $\hT = - \sfJ \hT \sfJ :=
\big( \begin{smallmatrix} a & 2b \\ 2b^* & -a \end{smallmatrix} \big) 
\in \cB[\fh \oplus \fh]$. 

Indeed, if we set 
$A_t^*(\tF) := e^{i t \QQ(T)} A^*\big( e^{-it\hT} \tF \big) e^{-i t \QQ(T)}$
for $t \in [0,1]$ then $\dot{A}_t^*(\tF) = 0$, by \eqref{eq-V.03}, and
hence $e^{i \QQ(T)} A^*\big( e^{-i\hT} \tF \big) e^{-i \QQ(T)}
= A_1^*(\tF) = A_0^*(\tF) = A^*(\tF)$ which directly yields \eqref{eq-V.04}
with $\tF := e^{i\hT} F$.

Note that we cannot directly quantize $\frac{1}{2}\hT$ in the sense of
\eqref{eq-V.02}, for if we replace $T$ by $\frac{1}{2}\hT$ in
\eqref{eq-V.02}, we obtain an expression $\frac{1}{2}\QQ(\hT)$, say,
which fulfilled $\frac{1}{2}\QQ(\hT) = \QQ(T) + \frac{1}{2}\Tr[a]$ and
would, hence, not exist in case that $a$ is not trace-class. Further
note that by the antilinearity of $\sfJ$, we have that 
$[i \hT] \sfJ = - i \sfJ T = \sfJ [i \hT]$ and hence 
$e^{-i\hT} \sfJ = \sfJ e^{-i\hT}$. Since $\QQ(T)$ is self-adjoint, 
$e^{i \QQ(T)} \in \Bog_\fF$ is a \GREEN{Bogoliubov} transformation with
\begin{align} \label{eq-V.05} 
\exp[i \QQ(T)] \ = \  \UU_{\exp[- i \hT ]} \, .
\end{align}
In fact, all \GREEN{Bogoliubov} transformations can be written in this form or,
at least, approximated in the strong topology. That is, we may
identify the \GREEN{Bogoliubov} transformations with the family of unitary
operators generated by $i$ times self-adjoint quadratic Hamiltonians,
\begin{align} \label{eq-V.06} 
\Bog_\fF \ = \ 
\ol{\Big\{ \exp[i \QQ(T)] \; \Big| \ 
T = \big( \begin{smallmatrix} a & b \\ b^* & 0 \end{smallmatrix} \big) 
\, , \ a \in \cB(\fh) \, , \ a \geq 0  \, , \ b \in \cL^2(\fh)
\Big\}} \, ,
\end{align}
where the bar denotes closure in the strong operator topology. 

\paragraph*{Quasifree Density Matrices:} It turns out that quadratic
Hamiltonians play an important role not only for \GREEN{Bogoliubov}
transformations, but also for density matrices. Recall from
\eqref{eq-IV.27} the definition of the density matrix 
$\rho_0 = \PP_1 Z_0^{-1} \exp[-\hh_0] \in \DM$, where
$\PP_1 = n_1 \, n_2 \cdots n_{K-1}$ and
\begin{align} \label{eq-V.07} 
\hh_0 \ = \
\sum_{\ell = K}^L \mu_\ell \: c^*(f_\ell) \, c(f_\ell)
\ = \  
\QQ(H_0) 
\end{align}
is the quadratic Hamiltonian corresponding to
\begin{align} \label{eq-V.08} 
H_0 \ := \ 
\begin{pmatrix} h_0 & 0 \\ 0 & 0 \end{pmatrix} 
\quad \text{and} \quad 
h_0 \ := \ \sum_{\ell =K}^L \mu_\ell \: |f_\ell \ra\la f_\ell| \, .
\end{align}
We now construct an approximation $\rho_\eps \in \QDM$ for 
$\rho_0$, such that $\rho_\eps \to \rho_0$ in $\DM_{\NN}$, as
$\eps \to 0$. 
For $\eps >0$, we define 
\begin{align} \label{eq-V.08,01} 
\tau_\ell(\eps) \ := \ 
\left\{ \begin{array}{cc} 
1 - \eps/K \, , & \ell < K \, , \\ 
\mu_\ell \, ,    & K \leq \ell \leq L \, , \\
\eps \, e^{-\ell^2} \, , & \ell > L \, , \\
\end{array} \right. 
\end{align}
noting that $\tau_\ell(\eps) \in (0,1)$, for all $\ell \in \ZZ^+$.
Next, we set $H_\eps := h_\eps \oplus 0$, where
$h_\eps := \ \sum_{\ell =1}^\infty \tau_\ell(\eps) \: |f_\ell \ra\la
f_\ell|$. Finally, $Z_\eps := \Tr_\fF\big( \exp[-\QQ(H_\eps)] \big)$ and 
$\rho_\eps := Z_\eps^{-1} \, \exp[-\QQ(H_\eps)]$.

We further recall from \eqref{eq-IV.30} that $\rho_0$, after
conjugation with the \GREEN{Bogoliubov} transformation $\UU_W \in \Bog_\fF$,
yields the density matrix $\trho_0 := \UU_W \rho_0 \UU_W^*$ whose
reduced generalized 1-pdm $\Gamma_{\trho_0}^{(1)}$ \RED{is equal to} the
prescribed generalized 1-pdm $\Gamma^{(1)}$ from
Theorem~\ref{thm-IV.01}. \BLUE{Therefore, if} $T_\eps = \big( 
\begin{smallmatrix} a_\eps & b_\eps \\ b_\eps^* & 0 \end{smallmatrix} \big)$,
$a_\eps \in \cB(\fh)$, $a_\eps \geq 0$, and $b_\eps \in \cL^2(\fh)$, is
such that $\UU_W \Psi = \lim_{\eps \to 0} \exp[i \QQ(T_\eps)] \Psi$, for all 
$\Psi \in \fF$, then in $\DM_{\NN}$
\begin{align} \label{eq-V.08,02} 
\trho_\eps \; := \; \exp[i \QQ(T_\eps)] \, \rho_\eps \, \exp[-i \QQ(T_\eps)]  
\ \to \ \trho_0 \qquad \eps \to 0 \, ,
\end{align}
due to the unitarity of $\UU_W$ and the fact that 
$\rho_0 \in \DM_{\NN}$. On the other hand,
\begin{align} \label{eq-V.09} 
\trho_\eps \ = \ 
Z^{-1} \exp[-\QQ(\tH_\eps)] \, , \quad \text{where} \quad
\tH_\eps \ = \ e^{-i\hT} \, H_\eps \, e^{i\hT} \, . 
\end{align}
So, defining the set of \RED{\textit{quasifree density matrices}}
\begin{align} \label{eq-V.10,01} 
\RED{\QDM} 
\ := \ 
\ol{\Big\{ Z^{-1} \, \exp[- \QQ(H_0)] \; \Big| \ 
H_0 \in \mathrm{qh} \, , \ Z := \exp[- \QQ(H_0)] < \infty \Big\}} \, ,
\end{align}
where the bar indicates closure in $\DM_{\NN}$ and
\begin{align} \label{eq-V.10,02} 
\mathrm{qh} \ := \ 
\left\{ \left. \begin{pmatrix} a & b \\ b^* & 0 \end{pmatrix}
\; \right| a = a^* \geq 0  \, , \ e^{-a} \in \cL^1(\fh) 
\, , \ b \in \cL^2(\fh) \right\} \, ,
\end{align}
we conclude from \eqref{eq-V.08,02}-\eqref{eq-V.09} that every 1-gpdm
is the 1-gRDM of a semigroup generated by a quadratic Hamiltonian or a
limit in $\DM_{\NN}$ thereof.

Note that $\QDM \subseteq \DM$. Further note that the closure in the
definition \eqref{eq-V.10,01} of quasifree density matrices is
important because otherwise the orthogonal projection 
\BLUE{$\rho = |\Phi_\uf \ra\la \Phi_\uf |$} onto the Slater determinant
\BLUE{$\Phi_\uf = f_1 \wedge \cdots \wedge f_N$} of orthonormal orbitals
$f_1, \ldots, f_N \in \fh$ would be excluded. The Slater determinant
\BLUE{$\Phi_\uf$}, however, is the \GREEN{Bogoliubov} transform $\UU_W \Om$ of the
vacuum vector, with $W = \big(
\begin{smallmatrix} P^\perp & \sfj P \sfj \\ P & \sfj P^\perp \sfj 
\end{smallmatrix} \big)$, where $P = \sum_{n=1}^N |f_n \ra\la f_n|$
is the orthogonal projection onto the subspace spanned by 
$f_1, \ldots, f_n \in \fh$. Hence, the corresponding rank-one
projection $\rho = |\UU_W \Om \ra\la \UU_W \Om |$ is a 
(pure) quasifree density matrix. The \GREEN{Bogoliubov} linear map
$W \in \Bog_{\fh \oplus \fh}$ is not of the form $\exp[-i \QQ(T)]$,
for any $T \in \mathrm{qh}$, but can be obtained as a strong limit
of these.

With the definitions in Eqs.~\eqref{eq-V.10,01}-\eqref{eq-V.10,02}, we
observe that the proof of Theorem~\ref{thm-IV.01} actually yields the
following stronger statement.
%
\begin{corollary} \label{cor-V.01} Let $\Gamma^{(1)} \in \genonepdm$
be a generalized one-particle density matrix. Then there exists
a unique quasifree density matrix $\eta \in \QDM$ such that 
$\Gamma^{(1)} = \Gamma_\eta^{(1)}$.
\end{corollary}
%
We \BLUE{do not} comment on the uniqueness part of Corollary~\ref{cor-V.01}
but point out the following important consequence.
%
\begin{corollary} \label{cor-V.02} The requirement 
$\Gamma_\rho^{(1)} = \Gamma_\eta^{(1)}$ defines a map 
\begin{align} \label{eq-V.11} 
q: \DM \ \to \ \QDM \, , \ \ \rho \mapsto q(\rho) = \eta \, .
\end{align}
For a density matrix $\rho \in \DM$, its image 
$\RED{q(\rho)} = \eta \in \QDM$ is
called its \RED{\textit{quasifree reduction}}.
\end{corollary}
%
For a quasifree state $\rho \in \QDM$ the reduced generalized 
$k$-pdm can be explicitly computed in terms of its reduced 
generalized $1$-pdm, as the following theorem asserts.
%
\begin{theorem} \label{thm-V.03} Let $\rho \in \DM$ be a density
matrix and denote $\la M \ra := \Tr_\fF[\rho M]$, for any 
$M \in \cB(\fF)$. Then the following statements are equivalent.
\begin{itemize}
\item[(i)] The density matrix $\rho \in \QDM$ is quasifree.

\item[(ii)] For all $k \geq 2$, all truncated $2k$-point functions
vanish, i.e., for all $F_1, F_2, \ldots, F_{2k} \in \fh \oplus \fh$,
\begin{align} \label{eq-V.12} 
\la A_1 \, A_2 \cdots A_{2k} \ra
\ = \ 
\sum_{\pi \in \cP_{2k}} (-1)^\pi \,  
\la A_{\pi(1)} A_{\pi(2)} \ra \, \la A_{\pi(3)} A_{\pi(4)} \ra 
\cdots \la A_{\pi(2k-1)} A_{\pi(2k)} \ra \, , 
\end{align}
where $A_i := A^*(F_i)$, $\cP_{2k}$ is the set of permutations
$\pi : \{1, 2, \ldots, 2k\} \to \{1, 2, \ldots, 2k\}$ that obey
$\pi(1) < \pi(3) < \ldots < \pi(2k-1)$ and $\pi(2j-1) < \pi(2j)$, 
for all $1 \leq j \leq k$, and $(-1)^\pi$ denotes its sign. 

\item[(iii)] All truncated \BLUE{four}-point functions
vanish, i.e., for all $F_1, F_2, F_3, F_4 \in \fh \oplus \fh$,
\begin{align} \label{eq-V.13} 
\la A_1 \, A_2 \, A_3 \, A_4 \ra
\ = \ 
\la A_1 A_2 \ra \la A_3 A_4 \ra 
- \la A_1 A_3 \ra \la A_2 A_4 \ra 
+ \la A_1 A_4 \ra \la A_2 A_3 \ra \, ,
\end{align}
where $A_i := A^*(F_i)$.
\end{itemize}
\end{theorem}
%
Characterization~(ii) of quasifree density matrices in the above
theorem is often taken as their definition. The somewhat surprising
statement for a given density matrix, that the \BLUE{sole} vanishing of
its truncated \BLUE{four}-point functions implies its quasifreeness
originates in a theorem of Marcinkiewicz \cite{Marcinkiewicz1939} in
(classical) probability theory. In the context of quantum physics, it
was first proved by Robinson \cite{Robinson1965a} and later
generalized in \cite{Baumann1975, BaumannHegerfeldt1985} for boson
systems. The generalization to fermions can be traced back to work of
Rajagopal and Sudarshan \cite{RajagopalSudarshan1974}, see also the
comment by Titulaer \cite{Titulaer1975}. We refer to Salmhofer
\cite{Salmhofer2009} for a modern presentation of truncated fermion
correlation functions.

The quasifree reduction $q: \DM \to \QDM$ defined in 
Corollary~\ref{cor-V.02} is a projection, i.e., an idempotent map 
$q^2 = q$ from the density matrices onto quasifree density matrices.
Gottlieb and Mauser \cite{GottliebMauser2007} observed that
the image $q(\rho) \in \QDM$ of $\rho \in \DM$ under this
projection is the closest element to $\rho$ in $\QDM$ in the sense
that it minimizes the relative entropy among all quasifree density
matrices, as the following Theorem asserts. 
%
\begin{theorem} \label{thm-V.04} Let $\rho \in \DM$ be a density
matrix and $q(\rho) \in \QDM$ its quasifree reduction. If
the relative entropy 
\begin{align} \label{eq-V.14} 
S[ \rho , q(\rho) ]
\ := \ 
\Tr_\fF\big\{ \rho \big( \log[\rho] - \log[q(\rho)] \big) \big\}
\ < \ \infty
\end{align}
exists, then 
\begin{align} \label{eq-V.15} 
S[ \rho , q(\rho) ]
\ = \ 
\inf_{\eta \in \QDM} \big\{ S[ \rho , \eta ] \big\} \, .
\end{align}
\begin{proof} Let $\eta \in \QDM$ be a quasifree density matrix
which, for simplicity, is assumed to be given as the exponential 
$\eta = Z^{-1} \, \exp[- \QQ]$
of a quadratic Hamiltonian $\QQ \equiv \QQ(H_0)$, for some
$H_0 \in \mathrm{qh}$, and that the von Neumann entropies 
$S[\rho] := - \Tr_\fF\{ \rho \: \log[\rho] \}$, $S[\eta]$, and 
$S[q(\rho)]$ of $\rho$, its quasifree reduction, and $\eta$ exist.
Then $-\log[q(\rho)] = \QQ + \log(Z)$ and hence
\begin{align} \label{eq-V.16} 
S[ \rho , \eta ]
\ = \ &
S[\rho] \, + \, \Tr_\fF\big\{ \rho \big( - \log[\eta] \big) \big\}
\ = \ 
\log(Z) \, + \, S[\rho] \, + \, 
\Tr_\fF\{ \rho \, \QQ \}
\\[1ex] \nonumber
\ = \ &
\log(Z) \, + \, S[\rho] \, + \, 
\Tr_\fF\{ q[\rho] \, \QQ \} 
\ = \ 
S[\rho] \, - \, \Tr_\fF\{ q[\rho] \, \log[\eta] \} \, ,
\end{align}
since quadratic observables have the same expectation value w.r.t.\ a
density matrix and its quasifree reduction. The same identity
holds true, if we replace $\eta$ by $q[\rho]$, and we obtain
\begin{align} \label{eq-V.17} 
S[ \rho , \eta ] - S[ \rho , q(\rho) ] 
\ = \ &
\Tr_\fF\big\{ q(\rho) \big( \log[q(\rho)] - \log[\eta] \big) \big\}
\ = \
S[ q(\rho) , \eta ] 
\ \geq \ 0 \, ,
\end{align}
since relative entropy is nonnegative.
\end{proof}
\end{theorem}
%
We \RED{note} that the existence of the von Neumann entropies 
$S[\rho] := - \Tr_\fF\{ \rho \: \log[\rho] \}$ of $\rho$ and 
$q[\rho]$ is assumed in the proof of Theorem~\ref{thm-V.04}
only for convenience and is not implied by the finiteness
of their relative entropy $S[\rho, q(\rho)]$. Note, however, that
if $\rho \in \QDM$ is quasifree and $S[\rho] < \infty$ then
it assumes the simple form 
\begin{align} \label{eq-V.18} 
S[\rho]
\ = \ & 
- \Tr_\fF\{ \rho \: \log[\rho] \}
\ = \ 
S^{(1)}[ \Gamma_\rho^{(1)} ] 
\ := \
- \Tr_{\fh \oplus \fh}\big\{ \Gamma_\rho^{(1)} \: 
\log[\Gamma_\rho^{(1)}] \big\} \, .
\end{align}
This is not hard to check for a quasifree density matrix 
$\rho_0 \in \QDM$ of the form $\rho_0 = \PP_1 \RRRED{Z_0}^{-1} \exp[-\hh_0]$, as
in \eqref{eq-IV.27} by explicit computation. The general identity
\eqref{eq-V.18} then follows from the invariance of $S[\rho]$ and
$S^{(1)}[ \Gamma^{(1)} ]$ under unitary transformations and the
application of a suitable \GREEN{Bogoliubov} linear map 
$W \in \Bog_{\fh \oplus \fh}$ to $\Gamma_\rho^{(1)}$ to transform
it to $\Gamma_{\rho_0}^{(1)}$ and the corresponding \GREEN{Bogoliubov} 
transformation $\UU_W \in \Bog_\fF$ to $\rho$ to transform
it to $\rho_0$. See also \cite{BachLiebSolovej1994}.

\newpage
\section{\BLUE{Bogoliubov--}Hartree-Fock Approximation and \\
Generalizations of Lieb's Variational Principle} 
\label{sec-VI}
%
\paragraph*{\BLUE{Bogoliubov--}Hartree--Fock Approximation:} Since quasifree
density matrices are, in particular, density matrices, we immediately
observe that the \BLUE{\textit{Bogoliubov--Hartree--Fock energy}}
\begin{align} \label{eq-VI.01}
\RED{E_\BHF}
\ := \  
\inf\big\{ \Tr_\fF( \rho \, \HH_\mu ) \; \big| 
\ \rho \: \in \: \QDM \, , \ \la \HH \ra_\rho < \infty \big\} 
\end{align}
defines an upper bound $E_\BHF \geq E_\gs$ on the total ground state
energy $E_\gs$ defined in \eqref{eq-II.12,02}. For a quasifree
density matrix $\rho \in \QDM$, the energy expectation value 
$\Tr_\fF( \rho \, \HH_\mu ) = \cE_\BHF(\Gamma_\rho^{(1)})$ 
depends only on its \BLUE{1-gRDM} 
$\Gamma_\rho^{(1)} = \big( \begin{smallmatrix} 
\gamma_\rho^{(1)} & \alpha_\rho \\ 
\alpha_\rho^* & \bfone - \sfj \gamma_\rho^{(1)} \sfj 
\end{smallmatrix} \big) \in \genonepdm$, where
\begin{align} \label{eq-VI.02}
\cE_\BHF(\Gamma_\rho^{(1)})
\ := \ &
\Tr_\fh\big[ h_\mu \, \gamma_\rho^{(1)} \big]
\: + \: 
\frac{1}{2} \Tr_{\fh \otimes \fh}\big[ V \, (\bfone - \Ex)
(\gamma_\rho^{(1)} \otimes \gamma_\rho^{(1)}) \big]  
\nonumber \\[1ex]
& \: + \: 
\frac{1}{2} \Tr_{\fh \otimes \fh}\big[ V \, \Ex \, 
(\alpha_\rho^* \otimes \alpha_\rho) \big] \, ,
\end{align}
where $h_\mu := h - \mu \bfone$. Moreover, since 
$\QDM \ni \rho \mapsto \Gamma_\rho^{(1)} \in \genonepdm$ is a bijection,
we obtain
\begin{align} \label{eq-VI.02,1}
E_\BHF \ = \  
\inf\Big\{ \cE_\BHF(\Gamma^{(1)}) \; \Big| 
\ \ \Gamma^{(1)} \, \in \, \genonepdm \Big\} \, . 
\end{align}
Note that if minimizers $\Gamma_\BHF^{(1)}$ exist then they necessarily
fulfill a stationarity condition, which in \BLUE{Bogoliubov--}Hartree--Fock
theory also takes the form of a self-consistent equation
\begin{align} \label{eq-VI.02,2}
\Gamma_\BHF^{(1)} 
\ = \ 
\bfone_\nonpos\big( h_\BHF[\Gamma_\BHF^{(1)}] \big) \, ,
\end{align}
where $h_\BHF[\Gamma^{(1)}]$ is again an effective Hamiltonian on $\fh
\oplus \fh$ and $\bfone_\nonpos( h_\BHF[\Gamma_\BHF^{(1)}] )$ is a
certain projection onto the eigenspaces of $h_\BHF[\Gamma_\BHF^{(1)}]$
of negative and zero eigenvalues such that $\bfone(
h_\BHF[\Gamma_\BHF^{(1)}] < 0 ) \leq \bfone_\nonpos(
h_\BHF[\Gamma_\BHF^{(1)}] ) \leq \bfone( h_\BHF[\Gamma_\BHF^{(1)}]
\leq 0 )$. The precise form of $\bfone_\nonpos$ is difficult to
determine because of the requirement $\bfone - \Gamma_\BHF^{(1)} =
\sfJ \Gamma_\BHF^{(1)} \sfJ$ which $\Gamma_\BHF^{(1)}$ and, therefore,
also $\bfone_\nonpos( h_\BHF[\Gamma_\BHF^{(1)}] )$ necessarily fulfills.

It is possible, however, to use \BLUE{a (further)} generalization of
\BLUE{Bogoliubov--}Hartree--Fock theory to positive temperatures
$1/\beta >0$, which is not reviewed here, and obtain a minimizer
$\Gamma_\BHF^{(1)}$ by the \BLUE{zero-temperature} limit $\beta \to
\infty$ of a family of minimizers $\big( \Gamma_\beta^{(1)}
\big)_{\beta \in \RR^+}$ for inverse temperature $1/\beta$. For fixed
$\beta$, the minimizer $\Gamma_\beta^{(1)}$ necessarily fulfills the
self-consistent equation
\begin{align} \label{eq-VI.02,3}
\Gamma_\beta^{(1)} 
\ = \ 
F_\beta\big( h_\beta[\Gamma_\beta^{(1)}] \big) \, ,
\end{align}
where $h_\beta[\Gamma]$ is a suitable effective Hamiltonian, itself
depending on $\beta$, and $F_\beta(x) = (1 + e^{\beta x})^{-1}$ is the
Fermi function. 

We remark that the \BLUE{Bogoliubov--}Hartree--Fock theory for
positive temperature derives from a variational principle,
namely the minimization of the \RED{\textit{Hartree--Fock pressure functional
$\RED{-\cP_\beta}$}} by
\begin{align} \label{eq-VI.02,4}
-\cP_\beta(\Gamma^{(1)})
\ := \  
\cE_\BHF(\Gamma^{(1)}) \: - \: \beta^{-1} \, S^{(1)}( \Gamma^{(1)} ) \, .
\end{align}
\RRRED{Lieb, Solovej, and the author} have demonstrated in
\cite{BachLiebSolovej1994} that it fulfills
\begin{align} \label{eq-VI.02,5}
\cP_\beta(\Gamma^{(1)})
\ \leq \,  
\beta^{-1} \, \log\big[ \Tr_\fF\big\{ \exp[-\beta \HH_\mu] \big\} \big] \, ,
\end{align}
for any generalized 1-pdm $\Gamma^{(1)} \in \genonepdm$ and, hence,
yields a lower bound to the pressure (in the sense of statistical
mechanics), in analogy to $\cE_\BHF(\Gamma^{(1)})$ being an
upper bound to the total ground state energy $E_\gs$.

\paragraph*{Repulsive Potentials:} If $\fh = L^2(M, d\nu)$ 
for a measure space $(M, d\nu)$, and 
$V(x,y) \geq 0$ is a repulsive potential, i.e.,
a nonnegative multiplication operator on $\fh \otimes \fh$, then
\begin{align} \label{eq-VI.03}
\Tr_{\fh \otimes \fh}\big[ V \, \Ex \, 
(\alpha_\rho^* \otimes \alpha_\rho) \big] 
\ = \
\int V(x,y) \, |\alpha_\rho(x,y)|^2 \: d\nu(x) \, d\nu(y)
\ \geq \ 0 \, .
\end{align}
In other words: For repulsive pair potentials the pairing operator
yields a nonnegative contribution to the energy. 
Now, if $\Gamma^{(1)} = \big( \begin{smallmatrix} 
\gamma^{(1)} & \alpha \\ \alpha^* & \bfone - \sfj \gamma^{(1)} \sfj 
\end{smallmatrix} \big) \in \genonepdm$ is a 
\RRRED{\sssout{generalized 1-pdm} 1-gpdm}, so is
$\tGamma^{(1)} :=  
\big( \begin{smallmatrix} \bfone & 0 \\ 0 & -\bfone \end{smallmatrix} \big)
\Gamma^{(1)}
\big( \begin{smallmatrix} \bfone & 0 \\ 0 & -\bfone \end{smallmatrix} \big)
= \big( \begin{smallmatrix} 
\gamma^{(1)} & -\alpha \\ -\alpha^* & \bfone - \sfj \gamma^{(1)} \sfj 
\end{smallmatrix} \big) \in \genonepdm$,
and by the convexity of $\genonepdm$ we conclude that 
$\hGamma^{(1)} = \frac{1}{2} \big( \Gamma^{(1)} + \tGamma^{(1)} \big) 
= \big( \begin{smallmatrix} 
\gamma^{(1)} & 0 \\ 0 & \bfone - \sfj \gamma^{(1)} \sfj 
\end{smallmatrix} \big)
\in \genonepdm$ is a  \RRRED{\sssout{generalized 1-pdm} 1-gpdm}, too. 
Its energy expectation 
value, however, is 
\begin{align} \label{eq-VI.05}
\cE_\BHF(\hGamma^{(1)})
\ = \ &
\Tr_\fh\big[ h_\mu \, \gamma^{(1)} \big]
\, + \, 
\frac{1}{2} \Tr_{\fh \otimes \fh}\big[ V \, (\bfone - \Ex)
(\gamma^{(1)} \otimes \gamma^{(1)}) \big]  
\nonumber \\[1ex] 
\ \leq \ &
\cE_\BHF(\Gamma^{(1)}) \, .
\end{align}
It follows \GREEN{that, for} repulsive pair potentials,
the \BLUE{Bogoliubov--}Hartree-Fock energy agrees with the
(total) Hartree-Fock energy and does not improve the approximation,
\begin{align} \label{eq-VI.06}
E_\BHF \ = \ E_\HF
\ = \  
\inf\big\{ \cE_\HF(\gamma) \; \big| 
\ \gamma \ \RRRED{\text{\sssout{$= \gamma^*$}}} \ \in \cL^1(\fh) \, , \ 
0 \leq \gamma \leq \bfone \big\} \, , 
\end{align}
where the \RED{\textit{total Hartree--Fock energy}} is defined as 
$\RED{E_\HF} := \inf_{N \in \ZZ^+} \BLUE{\{E_\HF(N)\}}$. We stress that
the total Hartree--Fock energy, as a function of the chemical potential
$\mu$, is the Legrende transform of the Hartree-Fock energy $E_\HF(N)$
for $N$ particles. 

\paragraph*{Attractive Potentials:} If $\fh = L^2(M, d\nu)$ and the
pair potential $V: M \times M \to \RR$ is strictly negative in some
subset of $M \times M$ then it may happen that 
$E_\BHF < E_\HF$\BLUE{, i.e.,} the \BLUE{Bogoliubov--}Hartree--Fock 
approximation is, indeed, better than
the original Hartree--Fock approximation and \RED{\textit{pairing}} occurs, 
i.e., all minimizers
$\Gamma^{(1)} = \big( \begin{smallmatrix} 
\gamma^{(1)} & \alpha \\ \alpha^* & \bfone - \sfj \gamma^{(1)} \sfj 
\end{smallmatrix} \big) \in \genonepdm$ have a nonvanishing pairing
operator $\alpha \neq 0$. In this case the Hartree--Fock equations
indicating the stationarity of the energy functional at the minimum
turn into BCS-type equations for which those found by \RRRED{Bardeen,
  Cooper, and Schrieffer in} \cite{BardeenCooperSchrieffer1957} for
the description of superconductivity are a special case. Because of
similarity, the stationarity condition is usually called \RED{the} 
\textit{BCS equation}, and the latter have been systematically
analyzed for \RED{translation-invariant} systems under the additional
assumption, or constraint, that only \RED{translation-invariant}
states enter the energy functional \cite{HainzlSeiringer2016}, see
Section~\ref{sec-VII}.

There is no general criterion for the occurence of \textit{pairing},
but in case the fermions in the model are electrons or other
spin-$\frac{1}{2}$ particles, the one-particle Hilbert space is of the
form $\fh = \hfh \otimes \CC^2$ with $\hfh = L^2(M, d\nu)$, the
interaction potential \RRRED{$V$} is spin-independent and purely
attractive, \RRRED{$V \leq 0$}, and the operators $h = \hat{h} \otimes
\bfone$ and \RRRED{$V = (-\hV) \otimes (\bfone \otimes \bfone)$} are
\textit{real}, i.e., $\sfj = \hsfj \otimes \bfone$, \RRRED{$\hsfj
  \hat{h} = \hat{h} \hsfj$}, and $(\hsfj \otimes \hsfj) \hV = \hV
(\hsfj \otimes \hsfj)$, an explicit characterization of pairing was
given by \RRRED{Fröhlich, Jonsson, and the author} in
\cite{BachFroehlichJonsson2009}: Under these assumptions, the energy
minimizing \RRRED{\sssout{generalized 1-pdm} 1-gpdm} always takes the form
\begin{align} \label{eq-VI.07}
\Gamma^{(1)} \ \equiv \ \Gamma^{(1)}[\hgamma] 
\ := \ 
\begin{pmatrix} 
\hgamma & 0 & 0 & \sqrt{\hgamma - \hgamma^2 \, }   \\ 
0 & \hgamma & - \sqrt{\hgamma - \hgamma^2 \, } & 0 \\ 
0 & - \sqrt{\hgamma - \hgamma^2 \, } & \bfone - \hgamma & 0 \\ 
\sqrt{\hgamma - \hgamma^2 \, } &  0  & 0 & \bfone - \hgamma \\ 
\end{pmatrix} \, ,
\end{align}
where the auxiliary 1-pdm $\hgamma \in \cL^1(\hfh)$, 
$0 \leq \hgamma \leq \bfone_\hfh$, on $\hfh$ minimizes the
resulting auxiliary functional
\begin{align} \label{eq-VI.08}
& \hcE_\aux(\hgamma) 
\ := \ 
\frac{1}{2}\cE_\BHF(\Gamma^{(1)}[\hgamma]) 
\ = \ 
\Tr_\hfh[\hat{h} \, \hgamma]  
\\[1ex] \nonumber
& - \frac{1}{2} \iint
\hV(x,y) \Big\{ \rho_\hgamma(x) \, \rho_\hgamma(y) 
- |\hgamma(x,y)|^2
+ \big| \sqrt{\hgamma - \hgamma^2}(x,y) \big|^2 \Big\} 
\: d\nu(x) \, d\nu(y) \, .
\end{align}
Note that the minimizer is real in the sense that 
$\hsfj \hgamma = \hgamma \hsfj$ and $\sfj \gamma = \gamma \sfj$. 
Further note that the pairing operator entering
$\Gamma^{(1)}[\hgamma]$ assumes the form
\begin{align} \label{eq-VI.08,01}
\alpha \ = \ 
\sqrt{\hgamma - \hgamma^2 \, } \otimes 
\begin{pmatrix} 
0 & 1 \\ 
- 1 & 0 \\ 
\end{pmatrix} \, ,
\end{align} 
where the second $2 \times 2$-matrix factor 
$\big( \begin{smallmatrix} 0 & 1 \\ -1 & 0 \\
\end{smallmatrix} \big)$ ensures the antisymmetry condition 
$\alpha^* = - \alpha = - \sfj \alpha \sfj$, which an admissible 
pairing operator necessarily fulfills according to 
\eqref{eq-IV.06,01}.

The physical system to which \cite{BachFroehlichJonsson2009} was
applied is a star consisting of neutrons, which are spin-$\frac{1}{2}$
fermions that attract each other by gravity. While it is generally
important to prove statements about minimization problems without
requiring the actual existence of a minimizer,
\cite{BachFroehlichJonsson2009} \BBBLUE{left} this existence question
unresolved. Lenzmann and Lewin, however, \BLUE{proved the existence of
  a minimizer for these neutron stars under natural conditions in
  \cite{LenzmannLewin2010}}.

\paragraph*{Dirac--Fock Equations:} Shortly after the discovery of
\textit{nonrelativistic} quantum mechanics and the formulation of the
Hartree--Fock approximation, a \textit{relativistic} analogue, the
\RED{\textit{Dirac--Fock (DF) equations}}, was formulated by Swirles
\cite{Swirles1935}. The proof of existence of solutions the
Dirac--Fock equations pose a considerably more difficult problem as
compared to proving this for the Hartree--Fock equations, due to the
unboundeness of the energy functional from below, direct methods from
the calculus of variations do not really apply, as was pointed out by
\RRRED{Chaix, Iracane, and Lions} in \cite{ChaixIracane1989,
  ChaixIracaneLions1989} who introduced and studied the
\GREEN{\textit{Bogoliubov--Dirac--Fock (BDF) model}}. An effective
renormalization and then control on the \RRRED{unboundedness} below of
the BDF model, i.e., of the Hartree--Fock energy functional for
electrons and positrons for small particle number and coupling
constant was first obtained by \RRRED{Barbaroux, Helffer, Siedentop, and
  the author} in \cite{BachBarbarouxHelfferSiedentop1999}. The
existence of solutions for Coulomb systems was shown by Esteban and
Séré \cite{EstebanSere1999} and by Paturel \cite{Paturel2000}. In the
case of atoms, \BLUE{Barbaroux, Farkas, Helffer, and Siedentop and
  Barbaroux, Esteban, and Séré} related the Dirac--Fock equations to
the Hartree-Fock equations of the electron-positron field in
\cite{BarbarouxFarkasHelfferSiedentop2005, BarbarouxEstebanSere2005}.
For the same model, \RRRED{Hainzl, Lewin, and Séré} proved in
\cite{HainzlLewinSere2005, HainzlLewinSere2009} the existence of a
minimizer and its uniqueness for the BDF model, but in contrast to
\cite{BachBarbarouxHelfferSiedentop1999} Hainzl et al.\ chose the
projection on the free Dirac sea as a reference and extended their
results later to atoms and molecules with small particle number and
small coupling constants. Huber and Siedentop proved
\cite{HuberSiedentop2007}, in turn, that the Dirac--Fock equations for
atoms \BBBLUE{possess} solutions if, among other smallness conditions on
coupling constants, the particle number $N$ is such that the shells of
the corresponding hydrogen-like Dirac operator are exactly filled.

\paragraph*{Generalization of Lieb's Variational Principle:} We come
back to Lieb's argument which establishes his variational
principle. The $N$-particle density matrix $\rho_\av$ in
\eqref{eq-IV.01} results from averaging the pure quasifree density
matrices $|\Phi(\ug^{(\theta)}) \ra\la \Phi(\ug^{(\theta)})|$
over all possible values of $\theta$. Hence, there exists at least
one choice of $\theta$ such that the energy expectation value
$\la \Phi(\ug^{(\theta)}) | \: \HH \Phi(\ug^{(\Theta)}) \ra$
of this pure quasifree density matrix is smaller or equal to the 
energy expectation $\Tr_\fF[\rho_\av \, \HH]$ of $\rho_\av$.

\RRRED{Derezinski, Napiorkowski, and Solovej}
\cite{DerezinskiNapiorkowskiSolovej2013} and, simultaneously,
\RRRED{Breteaux, Knoerr, Menge, and the author} generalized this
statement in \cite{BachBreteauxKnoerrMenge2014} and demonstrated that
the \BLUE{Bogoliubov--}Hartree--Fock energy $E_\BHF$ can be approximated by
energy expectation values of \BLUE{pure} quasifree density matrices
to arbitrary accuracy. Moreover, as we additionally point out here, if
$E_\BHF$ is a minimum then there is also a pure quasifree density
matrix among the minimizers.

The generalization does not only extend the set of density
matrices, over which the Hartree--Fock energy functional is being
varied, but also allows for any semibounded self-adjoint Hamiltonian
with no additional repulsiveness assumption on the pair potential
and not even on the form of the Hamiltonian, that it be a sum of
a one-body term and a pair interaction. 
%
\begin{theorem} \label{thm-VI.01} Suppose that $\HH_\mu = \HH_\mu^*$
is semibounded. Then $E_\BHF = \hE_\BHF$, where
\begin{align} \label{eq-VI.08,02}
\hE_\BHF
\ := \  
\inf\big\{ \la \Om \, | \; \UU^* \, \HH_\mu \, \UU \Om \ra
\; \big| \ \UU \in \Bog_\fF \big\} \, .
\end{align}
Moreover, if there is a quasifree density matrix 
$\rho_\BHF \in \QDM \cap \DM_{\NN}$ of finite particle number
such that $\Tr_\fF[ \rho_\BHF \HH_\mu ] = E_\BHF$, then there exists
a \GREEN{Bogoliubov} transformation $\UU_\BHF \in \Bog_\fF$ such that
$E_\BHF = \la \Om | \, \UU^* \HH_\mu \UU \Om \ra$. 
\begin{proof} We only prove the second part of the theorem and
assume that $\rho_\BHF \in \QDM \cap \DM_{\NN}$ is a quasifree 
density matrix of finite particle number expectation value and with
$\Tr_\fF[ \rho_\BHF \HH_\mu ] = E_\BHF$. The requirement of finiteness
of $\la \NN \ra_{\rho_\BHF}$ can be relaxed, but we do not carry this
out here. We can find a \GREEN{Bogoliubov} transformation $\tUU \in \Bog_\fF$
such that $\rho_\BHF$ takes the form $\rho_\BHF = \tUU^* \rho_0 \tUU$,
where $\rho_0 = \PP_1 Z^{-1} \exp[-\QQ(H_0)]$, with 
$\PP_1 = n_1 \, n_2 \cdots n_{K-1}$ and 
$\QQ(H_0) = \sum_{\ell = K}^L \mu_\ell \: c^*(f_\ell) \, c(f_\ell)$,
as in \eqref{eq-IV.27} and in \eqref{eq-V.07}, respectively.

Since $\HH_\mu$ is semibounded,
$\tHH := \HH_\mu - E_\gs \geq 0$, as a quadratic form, 
$\tE_\BHF := E_\BHF - E_\gs \geq 0$, and 
$\tE_\BHF = \Tr_\fF[ \rho_\BHF \tHH ] = 
\Tr_\fF[ \rho_\BHF^{1/2} \tHH \rho_\BHF^{1/2} ]$. It follows that
\begin{align} \label{eq-VI.09} 
\tE_\BHF \ = \ &
\Tr_\fF\big[ \rho_\BHF^{1/2} \, \tHH \, \rho_\BHF^{1/2} \big] 
\ = \ 
\sum_{\unu: |A(\unu)| < \infty} \big\la \rho_\BHF^{1/2} \tUU^* \BLUE{\Psi_\unu} 
\, \big| \: \tHH \, \rho_\BHF^{1/2} \tUU^* \BLUE{\Psi_\unu} \big\ra 
\\[1ex] \nonumber 
\ = \ &
\sum_{\unu: |A(\unu)| < \infty} \big\la \rho_0^{1/2} \BLUE{\Psi_\unu} 
\, \big| \: \tUU^* \tHH \tUU \, \rho_0^{1/2} \BLUE{\Psi_\unu} \big\ra
\ = \ 
\sum_{\unu \in \cA} \big\la \rho_0^{1/2} \BLUE{\Psi_\unu} 
\, \big| \: \tUU^* \tHH \tUU \, \rho_0^{1/2} \BLUE{\Psi_\unu} \big\ra \, ,
\end{align}
where the orthonormal basis 
$\big\{\BLUE{\Psi_\unu} \, \big| 
\unu \in \{0,1\}^{\ZZ^+} , \, |A(\unu)| < \infty \big\} \subseteq \fF$
is introduced in \eqref{eq-IV.28}, and the convergence of 
the series is guaranteed by the positivity of each term.
Moreover, the summation can be restricted to the subset 
$\cA := \{ \unu \in \{0,1\}^{\ZZ^+} | \: 
|A(\unu)| < \infty , \rho_0^{1/2} \BLUE{\Psi_\unu} \neq 0 \}$ 
of indices $\unu$, for which $\rho_0^{1/2} \BLUE{\Psi_\unu}$ is nonvanishing.
The latter vectors and the set $\cA$ can, however, be determined 
explicitly. Indeed, $\rho_0^{1/2} \BLUE{\Psi_\unu} \neq 0$ only if 
$A(\unu) \supseteq \{ 1, 2, \ldots, K-1 \}$, and in this case, 
up to a sign, we have that
\begin{align} \label{eq-VI.10} 
\rho_0^{1/2} \BLUE{\Psi_\unu} 
\ = \ &
c_1^* \cdots c_{K-1}^* \, \prod_{\ell \in A(\unu) \cap \{K, \ldots, L\}} 
\Big( \sqrt{ \lambda_\ell \, (1-\lambda_\ell)^{-1} \, } 
\, c_\ell^* \Big) \Om 
\nonumber \\[1ex] 
\ = \ &
\big\| \rho_0^{1/2} \BLUE{\Psi_\unu} \big\| \:
\prod_{\ell \in A(\unu) \cap \{1, \ldots, L\}} c_\ell^* \Om \, . 
\end{align}
It follows that $\cA := \big\{ \unu \in \{0,1\}^{\ZZ^+} \big| \: 
|A(\unu)| < \infty , A(\unu) \subseteq \{1, \ldots, L\} \big\}$
and that
\begin{align} \label{eq-VI.11} 
\tE_\BHF \ = \ &
\sum_{\unu \in \cA} \| \rho_0^{1/2} \BLUE{\Psi_\unu} \|^2 \:  
\la \tUU \BLUE{\Psi_\unu} \, | \: \tHH \tUU \BLUE{\Psi_\unu} \ra \, .
\end{align}
\RED{In addition, since} $\BLUE{\Psi_\unu}$ is a Slater determinant,
for each $\unu \in \cA$, the pure density matrix 
$|\tUU \BLUE{\Psi_\unu} \ra\la\tUU \BLUE{\Psi_\unu}| \in \QDM$ is
quasifree and, hence, 
$\la \tUU \BLUE{\Psi_\unu} | \tHH \tUU \BLUE{\Psi_\unu} \ra \geq
\tE_\BHF$.  Moreover, since 
$\| \rho_0^{1/2} \BLUE{\Psi_\unu} \|^2 >0$, for all $\unu \in \cA$ and
$\sum_{\unu \in \cA} \| \rho_0^{1/2} \BLUE{\Psi_\unu} \|^2 =
\Tr_\fF[\rho_0] = 1$, Eq.~\eqref{eq-VI.11} implies that
\begin{align} \label{eq-VI.12} 
\forall \, \unu \in \cA: \qquad 
\la \tUU \BLUE{\Psi_\unu} \, | \: \tHH \tUU \BLUE{\Psi_\unu} \ra 
\ = \ \tE_\BHF 
\end{align}
and thus the assertion.
\end{proof}
\end{theorem}
%

\newpage
\section{Symmetries and Restricted \\
Hartree--Fock Approximation} 
\label{sec-VII}
%
In this final section we discuss symmetries of the quantum system
under consideration. \BLUE{In some part, we follow the work of Lieb,
Solovej, and the author in \cite{BachLiebSolovej1994}, and we refer
the reader for more details to that paper.} 

We assume that the Hamiltonian $\HH = \hh + \frac{1}{2} \VV$ is given
in second quantized form as in \eqref{eq-II.08} with $\hh$ and $\VV$
as in \eqref{eq-II.18}-\eqref{eq-II.19} and to obey stability of
matter, i.e., that $\HH_\mu = \HH - \mu \NN$ is semibounded for
sufficiently small $\mu <0$ and hence $\HH_\mu + E_0 \geq 1$, for
sufficiently large $E_0 >0$.

A family $\cS$ of unitary operators $U \in \cS$ is called a
\RED{\textit{symmetry}} of \BLUE{$\HH_\mu$
\begin{align} \label{eq-VII.00,1}
: \Leftrightarrow \quad \forall \, U \in \cS: \quad 
U \, (\HH_\mu + E_0)^{-1} \ = \ (\HH_\mu + E_0)^{-1} \, U \, .
\end{align} 
}
Given a symmetry $\cS$, we define the 
\RED{\textit{restricted \BLUE{Bogoliubov--}Hartree--Fock (BHF) energy}} to be 
\begin{align} \label{eq-VII.01}
\RED{E_\BHF(\cS)} &
\ := \  
\\[1ex] \nonumber 
\inf & \big\{ \Tr_\fF( \rho \, \HH_\mu ) \; \big| 
\ \rho \: \in \: \QDM \, , \ \la \HH \ra_\rho < \infty \, , \ 
\forall \, U \in \cS: \ U \rho = \rho U \big\} \, .
\end{align}
Obviously, $E_\BHF(\cS) \geq E_\BHF$, and the approximation made by
the restricted BHF energy is not better, and potentially worse, than
the one without restriction. The importance of the restricted gHF
approxmiation, however, lies in its improved accessability to explicit
computation. Translation invariant generalized 1-pdm, for example, can
be diagonalized by Fourier transform, or rotationally invariant
generalized 1-pdm have a natural
decomposition in terms of spherical harmonics. \\[-2ex]
\begin{compactitem}
\item If $E_\BHF(\cS) = E_\BHF$ then the symmetry $\cS$ is called
  \RED{\textit{preserved}}.

\item If $E_\BHF(\cS) > E_\BHF$ then the symmetry $\cS$ is called
  \RED{\textit{broken}}.\\[-2ex]
\end{compactitem}
It turns out that both cases of preserved symmetry and broken symmetry
occur in different models. The reason is the hidden concavity of
Hartree--Fock functionals, which is used in the proof of
Theorem~\ref{thm-III.03} and which leads to instabilities at the
minimum of the restricted functional. We discuss symmetries on various
examples of physical interest.

\paragraph*{Closed Shell Theorem in Unrestricted Hartree--Fock Theory
  and Rotation of Atoms:} We first discuss rotation symmetry and come
back to the Hartree--Fock approximation as originally introduced for
atoms. The periodic table of the elements is usually described in
terms of angular momentum shells, which contain the electron
states. This picture implicitly assumes that the electron orbitals are
eigenfunctions of the angular momentum operators $L^2$ and
$L_z$. Indeed, the Hamiltonian and the Hartree--Fock functional of an
atom is invariant under rotations about the origin, where the atomic
nucleus is located. Its minimizers, however, do generally not
\BBBLUE{possess} this rotational symmetry unless we study the
restricted theory. Indeed, Griesemer and Hantsch show in
\cite{GriesemerHantsch2012} that \BLUE{the (unrestricted) HF minimizer 
becomes rotationally invariant for $N$ electrons that fill up the
lowest angular momentum shells (e.g., $N=2, 6, 10, 14, \ldots$), 
as $Z \gg N$ becomes sufficiently large. On the other hand,
it is not difficult to see that, without restriction by
symmetries, the HF minimizer of an atom with \BLUE{three} electrons, 
say, and a small nuclear charge $Z$ breaks rotational symmetry.}

This phenomenon is also reflected by the \textit{closed shell theorem}
in \cite{BachLiebLossSolovej1994}: If a HF minimizer 
$\gamma_\HF \in \onepdm$ for a Coulomb system of $N$ electrons exists, then 
it is the rank-$N$ orthogonal projection onto the smallest
$N$ eigenvalues $e_1 \leq e_2 \leq \ldots \leq e_N$ of
the corresponding effective Hamiltonian $h_\HF[\gamma_\HF]$,
as in \eqref{eq-III.23}, and the lowest spectral point of 
$h_\HF[\gamma_\HF]$ greater or equal \RRRED{\sssout{that} than} $e_N$ is 
\textit{strictly bigger} than $e_N$, 
\begin{align} \label{eq-VII.02}
e_{N+1} \ := \ 
\inf\GREEN{\Big\{} \sigma\big( h_\HF[\gamma_\HF] \big) 
\setminus \{ e_1, e_2, \ldots, e_N\} \Big\} 
\ > \ e_N \, .
\end{align}
The interpretation of this statement is that, in Hartree--Fock
approximation, atoms and molecules never possess an open shell because
the highest energy level is always fully occupied. In particular, rare
earth elements with one loosely bound valence electron in a degenerate
high momentum shell do no occur in Hartree--Fock theory. Therefore,
the Hartree--Fock approximation for a single Lithium atom, say, does
not yield orbitals which are products of \BLUE{a radial function, a
spherical harmonic, and a spinor}.

\paragraph*{Particle Number Conservation:} The strongly continuous
one-parameter group $\cN = \big( \exp[-it \NN] \big)_{t \in \RR}$ of
unitary operators generated by the particle number operator $\NN$ is a
symmetry of all Hamiltonians $\HH$ of the form \eqref{eq-II.08}, as
these conserve particle number.

In Section~\ref{sec-VI} it is demonstrated that in case of a repulsive
potential, choosing a vanishing pairing operator is always favorable
for the energy minimization, and the particle number symmetry is
always preserved. For attractive potentials, this is not always the
case and, depending on the model, the particle number symmetry 
\BLUE{is preserved in some cases and broken in others}.

\paragraph*{Translation Invariance in $\RR^3$:} Three-dimensional
systems are translation invariant, if (the resolvent of) $\HH$
commutes with all $U_\va \in \cT_{\RR^3}$, where
$\cT_{\RR^3} = \{ U_\va | \va \in \RR^3 \}$ and 
$U_\va = \exp[ - i \va \cdot \vpp]$ is the translation by 
$\va \in \RR^3$. These translations are generated by 
$\vpp = (\bbp_1, \bbp_2, \bbp_3)$, where 
$\bbp_\nu = \sum_{j,k=1}^\infty \la f_j | (-i \partial_\nu) \ra
c^*(f_j) c(f_k)$ is the second quantization of the momentum operator
$-i \partial_\nu$ in the $\nu^{th}$ coordinate direction.  

From a physics point of view, it would be desirable to define these
translational invariant systems with the (single-fermion)
configuration space $\RR^3$ as described above\RRRED{. This} would
necessitate \BBBLUE{general} states, rather than density matrices, and
ultimately require an operator algebraic framework which we cannot
provide here.

\paragraph*{Translation Invariance on a large Torus:} To circumvent
the problem related to the thermodynamic (i.e., infinite volume) limit
it is customary to replace the configuration space $\RR^3$ by a torus
$\Lambda := (\RR / L \ZZ)^3$ of large, but finite, sidelength $L \gg
1$. The Hamiltonian $\HH$ then commutes with translations $U_\va =
\exp[ - i \va \cdot \vpp]$ by $\va \in \Lambda$ modulo $L$ in
$\cT_\Lambda := \{ U_\va | \va \in \Lambda \}$. The resulting model is
called \textit{Fermi Jellium} or \textit{Fermi gas}, and one is
interested in the limit $L \to \infty$ and in the energy per unit
volume $e_\gs := \lim_{L \to \infty} \{ L^{-3} \, E_\gs \}$\RRRED{. As}
the ground state energy \RRRED{(at fixed $\mu$)} is an
\textit{extensive} quantity, \RRRED{\sssout{and}} so are the 
\BLUE{Bogoliubov--}Hartree--Fock energy $E_\BHF$ and the restricted
\BLUE{Bogoliubov--}Hartree--Fock energy $E_\BHF(\cT_\Lambda)$. 
For this reason, we define the respective energies 
$e_\BHF := \lim_{L \to \infty} \{ L^{-3} \, E_\BHF \}$ and
$e_\BHF(\cT_\Lambda) := \lim_{L \to \infty} \{ L^{-3} \,
E_\BHF(\cT_\Lambda) \}$ per unit volume.

More than fifty years ago, Overhauser considered the above model with
\BBBLUE{a repulsive} interaction, for which \RRRED{the} pairing operator
vanishes and the \BLUE{Bogoliubov--}Hartree--Fock energy agrees with the
original total Hartree--Fock energy. At high density, the 
\BLUE{\textit{paramagnetic state}} represented by a Slater determinant
of plane waves occupying for both spin-up and spin-down electrons all
momenta $k \in \Lambda^*$ below the Fermi energy, i.e., for which
\BLUE{the dispersion $\om(k) \leq \mu$ is below the chemical
  potential}, is the natural tranlation invariant HF minimizer and
yields $\BLUE{e_\para :=} e_\BHF(\cT_\Lambda)$. He demonstrated in
\cite{Overhauser1960, Overhauser1962, Overhauser1968}, however, that a
lower energy $e_\BHF < e_\BHF(\cT_\Lambda)$ is produced by Slater
determinants which are not translation invariant but represent a spin
wave. The precise Hartree--Fock minimizer breaking the translation
invariance is not known explicitly, but in a recent paper
\cite{GontierHainzlLewin2019} \RRRED{Gontier, Hainzl, and Lewin}
estimated the difference $e_\BHF(\cT_\Lambda) - e_\BHF >0$ of the
energies and proved that it is exponentially small in the interaction
coupling. Thus, although the restricted HF energy is higher than the
HF energy without restriction, the two terms agree to any order in
powers of the coupling constant.

\paragraph*{The BCS Model - Spin Invariance:} 
We further introduce global spin transformations which rotate the spin
variables $\CC^2$ at each point in space by the \textit{same} unitary
\BLUE{transformation $u \in SU(2)$}. As the Hamiltonian is invariant under
such global spin rotations, this defines an additional symmetry
$SU(2)$ of the system. (One variant of) The \RED{\textit{BCS model}} is now
defined to be the restricted \BLUE{Bogoliubov--}Hartree--Fock energy
$E_\BCS(\cT_\Lambda \times SU(2))$. In the simplest model case,
$E_\BCS(\cT_\Lambda \times SU(2))$ can be explicitly computed thanks
to the restriction of the variation to translation-invariant
generalized 1-pdm which are in the same spin singlet state at any
point in $\Lambda$. Additionally choosing $\sfj$ to be complex
conjugation in Fourier space, the \BLUE{Bogoliubov--}Hartree--Fock
energy functional is varied only over 
$\Gamma^{(1)} \in \genonepdm$ of the form
\begin{align} \label{eq-VII.03}
\Gamma^{(1)}(k, k') 
\ = \ 
\delta_{k,k'} \: \begin{pmatrix}
\hgamma(k) \otimes 
\big( \begin{smallmatrix} 1 & 0 \\ 0 & 1 \\ \end{smallmatrix} \big) 
&
\halpha(k) \otimes 
\big( \begin{smallmatrix} 0 & 1 \\ -1 & 0 \\ \end{smallmatrix} \big) 
\\[1ex]
\halpha(k)^* \otimes 
\big( \begin{smallmatrix} 0 & 1 \\ -1 & 0 \\ \end{smallmatrix} \big) 
& 
[\bfone - \hgamma(k)] \otimes 
\big( \begin{smallmatrix} 1 & 0 \\ 0 & 1 \\ \end{smallmatrix} \big) 
\end{pmatrix} \, , 
\end{align}
where $\hgamma \in L^1(\Lambda^*; \RR_0^+)$ and 
$\halpha \in L^2(\Lambda^*)$ and $\Lambda^* = \frac{2\pi}{L} \ZZ^3$.
Inserting this into the energy functional at zero temperature gives
\begin{align} \label{eq-VII.04}
\cE_\BCS(\Gamma^{(1)})
\ = \ &
\sum_{k \in \Lambda^*} \big( \om(k) - \mu \big) \, \hgamma(k)
\: + \: 
\frac{1}{2} \|\hV\|_1 \|\hgamma\|_1^2 
\\[1ex] \nonumber
& \: - \: 
\frac{1}{2} \int_\Lambda V(x) \, |\gamma(x)|^2 \: d^3x
\: + \: 
\frac{1}{2} \int_\Lambda V(x) \, |\alpha(x)|^2 \: d^3x \, ,
\end{align}
where $V$, $\gamma$, and $\alpha$ are the inverse Fourier transform of
$\hV$, $\hgamma$, and $\halpha$, respectively. The potential $V$ is
assumed to be negative (attractive) for some part of $\Lambda$ in
order not to rule out nonvanishing $\alpha \neq 0$ to begin with. This
model and its variants, questions of existence and uniqueness of its
minimizers, the characterization of the resulting minimizers by the
BCS gap equation and the analysis of its solution for \RRRED{zero}
    and positive temperatures have been analyzed and physically
interpreted by Hainzl and Seiringer and others in a remarkable series
of papers \cite{HainzlHamzaSeiringerSolovej2008,
  FrankHainzlNabokoSeiringer2007, HainzlSeiringer2008a,
  HainzlSeiringer2008b, FreijiHainzlSeiringer2012,
  BraeunlichHainzlSeiringer2014, BraeunlichHainzlSeiringer2016}, see
\cite{HainzlSeiringer2016} for a review. 
\BBBLUE{Spin symmetry breaking and a phase transition between a 
ferromagentic and a paramagnetic phase has been recently proved
for a Hartree--Fock model like \eqref{eq-VII.04} under the
additional assumption of the absence $\alpha = 0$ of pairing,
i.e., restriction to conserved particle numbers, by Gontier and
Lewin in \cite{GontierLewin2019}.}

\BLUE{\paragraph*{The Hubbard Model:} 
The Hubbard model is a translation and spin invariant model on a
finite-dimensional one-particle Hilbert space $\fh = \ell^2(\Lambda)
\otimes \CC^2$. We describe the most frequently studied case when
$\Lambda := (\ZZ/L\ZZ)^d$ is the hypercubic $d$-dimensional periodic
lattice and the kinetic energy is represented by a real
nearest-neighbour hopping matrix $T = t \otimes \bfone$, where $t(x,y)
= -1$ if $x, y \in \Lambda$ are neighbouring lattice sites, and
$t(x,y) = 0$ otherwise. The pair potential $V(x-y) = \delta_{x,y}$ is
on-site only and coupled in by a coupling constant $\lambda \geq 0$ 
which we here assume to be positive, so that BHF minimizers and HF 
minimizers agree. In spite of its simplicity, the physical properties
of the Hubbard model change dramatically, as the model parameters
$\mu$ and $\lambda$ vary.}

\BLUE{ 
A special model situation in which the Hubbard model possesses a
rather large symmetry group is given at \textit{half-filling}, when
$\mu$ is chosen so that the density in the ground state equals $1$,
i.e., the number of lattice sites, $N = |\Lambda|$. At half-filling,
the HF minimizer was explicitly determined by Lieb, Solovej, and the
author in \cite{BachLiebSolovej1994}. It turns out that both the
translation symmetry and the spin symmetry are broken in this case:
The HF minimizers $\gamma_\HF$ exhibit antiferromagnetic order, i.e.,
$\vv_\HF(x) := 
\Tr_{\CC^2}\big[ \gamma_\HF(x,x) \, \vsigma \big] = (-1)^x \Delta
\vec{e}$, where $\vsigma = (\sigma_x, \sigma_y, \sigma_z) \in \CC^{2
  \times 2}$ are the Pauli matrices, $(-1)^x$ is an alternating sign
taking the value $1$ on even sites and $-1$ on odd sites, $\vec{e} \in
\RR^3$ is a(n arbitrary) unit vector, and $\Delta > 0$ is a
self-consistent gap parameter - much like the gap in BCS theory.}

\BLUE{
For a different choice of the chemical potential $\mu$ such that the
density $0 < \rho = N / |\Lambda| \GREEN{\ll} 1$ is small and the interaction
coupling $\lambda \gg 1$ is large, the Hartree-Fock approximation
restricted by the symmetry group $\cS_z$ generated by the spin-z
operator $S_z = \sum_{x \in \Lambda} \sigma_z$ was analyzed by Lieb,
Travaglia, and the author in \cite{BachLiebTravaglia2006}. Here,
the HF minimizer shows maximal ferromagnetic ordering in
the sense that $\vv_\HF(x) = \pm \eta \vec{e}_3$ with $\eta > 0$
taking its maximally possible value. In other words, in
the minimizing 1-pdm either all
electrons have spin up or all
electrons have spin down. In particular, the paramagnetic state
is not energetically favorable in this situation.}

\paragraph*{Periodic structures:} If the configuration space $\RR^3$
is again a torus $\Lambda_L := (\RR / L \ZZ)^3$ for some large integer
$L \gg 1$ then the Hamiltonian often commutes only with integer
translations $\va$ contained in the subgroup $\ZZ_L^3 = (\ZZ / L
\ZZ)^3 \subset \Lambda$, leading to the symmetry $\cS_{\ZZ_L^3}$ of
$\HH$. A typical example is a system of the form $\HH = \hh +
\frac{1}{2} \VV$ with $\VV$ having the full translation symmetry
$\cS_\Lambda$ but $\hh$ having only the smaller symmetry
$\cS_{\ZZ_L^3}$ due to the presence of a periodic external potential.

The closed shell theorem described above does not only hold for
Coulomb systems, but for general $N$ fermion systems with a repulsive
interaction potential (for which the generalized and the original
Hartree--Fock approximation coincide). In general, it may fail,
however, in case of restricted Hartree--Fock minimizers. For periodic
systems, the existence of minmizer was established by \RRRED{Catto, Le
  Bris, and Lions} in \cite{CattoLeBrisLions2001} where the 1-pdm are
restricted \RRRED{to} those which are invariant under (integral) lattice
translations. Ghimenti and Lewin have later shown in
\cite{GhimentiLewin2009a} a kind of closed shell theorem and proved
that the minimizer is a projection onto the smallest energies of the
corresponding Hartree--Fock effective operator.

\paragraph*{Acknowledgement:} I thank Mathieu Lewin and Konstantin
Merz for numerous very helpful corrections and suggestions.
\BLUE{Furthermore, I thank IPAM at UCLA, where part of this work was
  done, for its great hospitality.}


\end{document}